\documentclass[12pt,a4paper]{article}
\usepackage{hyperref}
\usepackage[T2A]{fontenc}
\usepackage[cp1251]{inputenc}
\usepackage[english]{babel}
\usepackage{amsmath,amssymb,amsfonts}
\usepackage{bm,latexsym,euscript,textcomp}
\usepackage{graphicx,color,graphics,caption,epsf,dcolumn}
\usepackage{natbib}
\usepackage{gensymb}
\usepackage{epstopdf}
\usepackage{booktabs}

\newcommand{\beq}{\begin{equation}}
\newcommand{\eeq}{\end{equation}}

\begin{document}

\title{\Large \bf  Heat engines and heat pumps\\ in a hydrostatic atmosphere:\\
How surface pressure and temperature constrain\\ wind power output and circulation cell size}

\author{A. M. Makarieva$^{1,2}$\thanks{\textit{Corresponding author.} {E-mail: ammakarieva@gmail.com}}, V. G. Gorshkov$^{1,2}$,
A.V. Nefiodov$^1$,\\ D.  Sheil$^{3}$, A. D. Nobre$^4$, P. Shearman$^{5,6}$, B.-L. Li$^2$}

\date{\vspace{-5ex}}

\maketitle

\noindent
$^{1}$Theoretical Physics Division, Petersburg Nuclear Physics Institute,
188300 Gatchina, St.~Petersburg, Russia,
$^{2}$XIEG-UCR International Center for Arid Land Ecology, University of California, Riverside 92521-0124, USA,
$^{3}$Norwegian University of Life Sciences, \AA s, Norway,
$^{4}$Centro de Ci\^{e}ncia do Sistema Terrestre INPE, S\~{a}o Jos\'{e} dos Campos SP 12227-010, Brazil
$^{5}$UPNG Remote Sensing Centre,  Biology Department, University of Papua New Guinea, Papua New Guinea,
$^{6}$School of Botany and Zoology, The Australian National University, Canberra, Australia.

\begin{abstract}
The kinetic energy budget of the atmosphere's meridional circulation cells is analytically assessed.
In the upper atmosphere kinetic energy generation grows with increasing surface temperature difference $\Delta T_s$ between the cold and
warm ends of a circulation cell; in the lower atmosphere it declines. A requirement that kinetic energy generation is positive in
the lower atmosphere limits the poleward cell extension $L$ of Hadley cells via a relationship between $\Delta T_s$ and
surface pressure difference $\Delta p_s$: an upper limit exists when $\Delta p_s$ does not grow with increasing $\Delta T_s$. This pattern is
demonstrated here using monthly data from MERRA re-analysis. Kinetic energy generation along air streamlines
in the boundary layer does not exceed $40$~J~mol$^{-1}$; it declines with growing $L$ and reaches zero for the largest
observed $L$ at 2~km height.  The limited meridional cell size necessitates the appearance of heat pumps --
circulation cells with negative work output where the low-level air moves towards colder areas. These cells
consume the positive work output of the heat engines -- cells where the low-level air moves towards the warmer areas --
and can in theory drive the global efficiency of atmospheric circulation down to zero. Relative contributions
of $\Delta p_s$ and $\Delta T_s$ to kinetic energy generation are evaluated: $\Delta T_s$ dominates in the upper atmosphere,
while $\Delta p_s$ dominates in the lower. Analysis and empirical evidence indicate that the net kinetic power output on
Earth is dominated by surface pressure gradients, with minor net kinetic energy generation in the upper atmosphere.
The role of condensation in generating surface pressure gradients is discussed.
\end{abstract}

\section{Introduction}

\noindent
The Earth has three meridional circulation cells in each hemisphere: Polar, Ferrel and Hadley.
In the first and last the air moves from the cold to the warm areas in the lower atmosphere
returning in the upper atmosphere, while in the intermediate cell the relationship between the air motion and temperature
gradient is reversed. Why do we find this pattern of cells rather than say
one cell in each hemisphere stretching from the equator to the pole?
What determines the size of the existing cells? Besides their theoretical significance, these questions
have practical implications \citep{webster04}. Shifting cell boundaries can lead to marked climatic
changes such as the decrease of rainfall observed in the subtropics \citep[e.g.,][]{bony15,heffernan16}.

On a rotating planet the maximum poleward extension of the meridional circulation cells in the upper atmosphere
is related to conservation of angular momentum. As the air moves poleward and approaches the Earth's rotation axis, the air velocity must
increase provided angular momentum is conserved. In the limit, at the pole, the air would need to reach an infinite ve\-lo\-ci\-ty.
The energy for this acceleration is derived from the pressure gradient in the upper troposphere,
which is associated with meridional temperature gradient. At a given height in the upper troposphere, air pressure above the warm areas is higher than it is over cold areas. The greater the temperature gradient, the further it can push the air towards the pole.
Thus, for a meridional flow conserving angular momentum the maximum cell size grows with increasing temperature difference between the equator and the pole.

If the angular momentum is not conserved but decreases as the air moves from the equator to the pole,
the air velocity increases more slowly. Then, for the same temperature difference, the meridional cell can reach further towards the pole.
The degree to which angular momentum is conserved is controlled by turbulent friction,
which determines the exchange of angular momentum between different atmospheric layers and the Earth.
Formally, in global circulation models turbulent friction is governed by several parameters like surface drag,
eddy diffusivity and orography. The effect of turbulent friction on the extension of circulation cells has received considerable attention from different perspectives \citep{held80,robinson97,schneider06,chen07,marvel13}.
For example, in the absence of orography if eddy diffusivity is exactly zero, the meridional cells do not exist. If, on the other hand,
turbulent friction in the upper atmosphere is sufficiently large, then the meridional circulation cells can extend to the poles
\citep[see, e.g.,][their Fig.~3]{marvel13}.

Provided there is sufficient information on turblent friction, the extension and intensity of the meridional circulation cells can be retrieved by solving the equations of motion. However, turbulent processes themselves depend on the nature and intensity of the large-scale circulation that they help to arrange. Furthermore, turbulent eddies on different scales can have their local energy sources rather than being simply products of dissipation of the large-scale motions. Since the general theory of atmospheric turbulence is absent,
model parameters that govern turbulent friction in circulation models cannot be specified a priori. Instead, they are chosen such
that the resulting circulation conforms to observations. However, given the crucial role
of the general circulation for the planetary climate (see, e.g., \citet{bates12}
and \citet{shepherd14} for two complementary perspectives), the fact that
turbulent processes cannot be formulated from theory hampers reliable predictions of future climates.

In this situation it is relevant to search for physical constraints to which the atmosphere
obeys which could govern formulation of the dissipative processes in the atmosphere.
In this paper we consider the process of kinetic energy generation in the upper and lower
atmosphere and how it can be expressed via the values of surface pressure and temperature.

Why is kinetic energy generation relevant and how does it relate to turbulence?
In a stationary atmosphere the rate of kinetic energy generation is equal to the rate at which wind power dissipates
via turbulent processes. However, unlike turbulence, kinetic energy generation can be easily formulated in terms of measurable atmospheric variables. Kinetic energy is generated by pressure gradients: wind power (the rate of kinetic energy generation) per unit air volume is equal to the scalar product of air velocity and pressure gradient \citep[e.g.,][]{boville03}.
Consider for simplicity an axisymmetric atmosphere. If eddy diffusivity is zero, the air is in geostrophic equilibrium and rotates around the Earth along the isobars everywhere except at the surface where wind velocity is zero. Since in
geostrophic balance air velocity is everywhere perpendicular to pressure gradient, the wind power is zero. No heat engines and no thermodynamic cycles exist on such an Earth. This means that turbulence not only controls the non-conservation of angular momentum along the streamlines of a large-scale circulation. It also ultimately determines whether there is a heat engine operating on Earth and how much power it generates.

The concept of Carnot heat engine as applied to the atmosphere has been considered by many authors
attempting to constrain the efficiency of atmospheric circulation \citep{wulf52,pe92,lorenz02,pa00,pauluis11,kieu15}.
The heat source is the warm Earth's surface and the heat sink is associated with a higher
altitude and/or latitude. Despite the prevalence of the Carnot cycle concept in the meteorological literature
no established relationships show how work output depends on pressure and temperature differences at the Earth's surface.
Here we develop such relationships for a Carnot cycle (Section 3) and for a more realistic cycle where the non-vertical streamlines go parallel to the surface at a constant height (Section 4). We show that kinetic energy generation along air streamlines in the lower atmosphere and in the upper atmosphere have opposite relationships on surface temperature difference. The greater the temperature difference between the warmer and the colder ends of the circulation in a heat engine, the more kinetic energy is generated in the upper atmosphere and less in the lower.

We then show that the condition that the kinetic energy generation must be positive
in the lower atmosphere limits the horizontal dimension of the cell via a relationship between $\Delta p_s$ and $\Delta T_s$.
An upper limit exists when $\Delta p_s$ does not grow with increasing $\Delta T_s$.
Using data from MERRA re-analysis \citep{rien11} we investigate the dependence between $\Delta p_s$ and $\Delta T_s$ for Hadley cells to find that this pattern matches the observations (Section 5).

In Section 6 we use the developed expressions to evaluate the kinetic energy generation budget
in the atmosphere composed of several heat engines (cycles with positive work output, Hadley and Polar cells)
and heat pumps (cycles with negative work output, Ferrel cells). We show that our theoretical estimates
are consistent with direct assessments of kinetic energy generation
by the Earth's meridional circulation cells (Section 7). We then consider the high-resolution estimates of global rates of kinetic energy
generation by \citet{huang15} and find that the net kinetic energy generation in the upper atmosphere is relatively small.
We infer that on Earth the surface pressure gradients are the principle determinants of the global kinetic power budget.

\section{Kinetic energy generation and total work output in a thermodynamic cycle}

In a hydrostatic atmosphere kinetic energy is mostly generated by horizontal pressure gradients.
On the other hand, major pressure differences are associated with the vertical dimension.
We start our analysis by considering how kinetic energy generation relates to the total work $A$ of a
thermodynamic cycle.

We use the ideal gas law
\begin{equation}
\label{ig}
pV = RT, \quad pdV = RdT - Vdp,
\end{equation}
where $R=8.3$~J~mol$^{-1}$~K$^{-1}$ is the universal gas constant, $p$ is pressure, $T$ is temperature and $V$ (m$^3$~mol$^{-1}$) is molar volume,
and the condition of hydrostatic equilibrium
\begin{equation}\label{he}
\rho \mathbf{g} = \nabla_z p,
\end{equation}
where $\rho = M/V$ is air density, $M$ is air molar mass and $\mathbf{g}$ is the vector of gravity acceleration.

Using the definition of horizontal velocity $u_h \equiv dy/dt$ and vertical velocity $w \equiv dz/dt$ we find
\begin{equation} \label{Wtotg}
A \equiv \oint pdV = -\oint V\frac{dp}{dt} dt = -\oint \left.V\frac{\partial p}{\partial y}\right|_{z=z(y)} dy,
\end{equation}
as far as from (\ref{he}) we have
\begin{equation}
-\oint \left.V\frac{\partial p}{\partial z}\right|_{y=y(z)} dz = \oint V \rho g dz = Mg \oint dz = 0.
\end{equation}
Here $y=y(z)$ and $z=z(y)$ are the equations of the closed streamline that defines the cycle, where $z$ is height and $y$ is horizontal coordinate.
Eq.~(\ref{Wtotg}) shows that for any thermodynamic cycle where the mass of gas is constant, total work $A$ (J~mol$^{-1}$) is equal to total
kinetic energy generation -- and this depends solely on horizontal pressure gradients. In an atmosphere where water vapor undergoes phase transitions,
Eq.~(\ref{Wtotg}) still describes kinetic energy generation per mole circulating air with
reasonable accuracy (see \ref{apA} for details).

We emphasize that the obtained expression (\ref{Wtotg}) for kinetic energy generation
do not carry any information about planet rotation and universally apply to any air streamlines, on
rotating as well as on non-rotating planets. This is because the Coriolis force that distinguishes the equations of motions
on a rotating planet from non-rotating one is by definition perpendicular to air velocity. In the result, it makes
no contribution to wind power (kinetic energy generation per unit time). Thus, while wind {\it velocity} can be approximately
retrieved from the assumption of geostrophic balance, wind {\it power} cannot. Wind power is non-zero to the degree by which the
geostrophic balance is broken.

\section{Kinetic energy generation and cell size limit in a Carnot cycle}

\begin{figure*}
\centerline{
\includegraphics[width=0.97\textwidth,angle=0,clip]{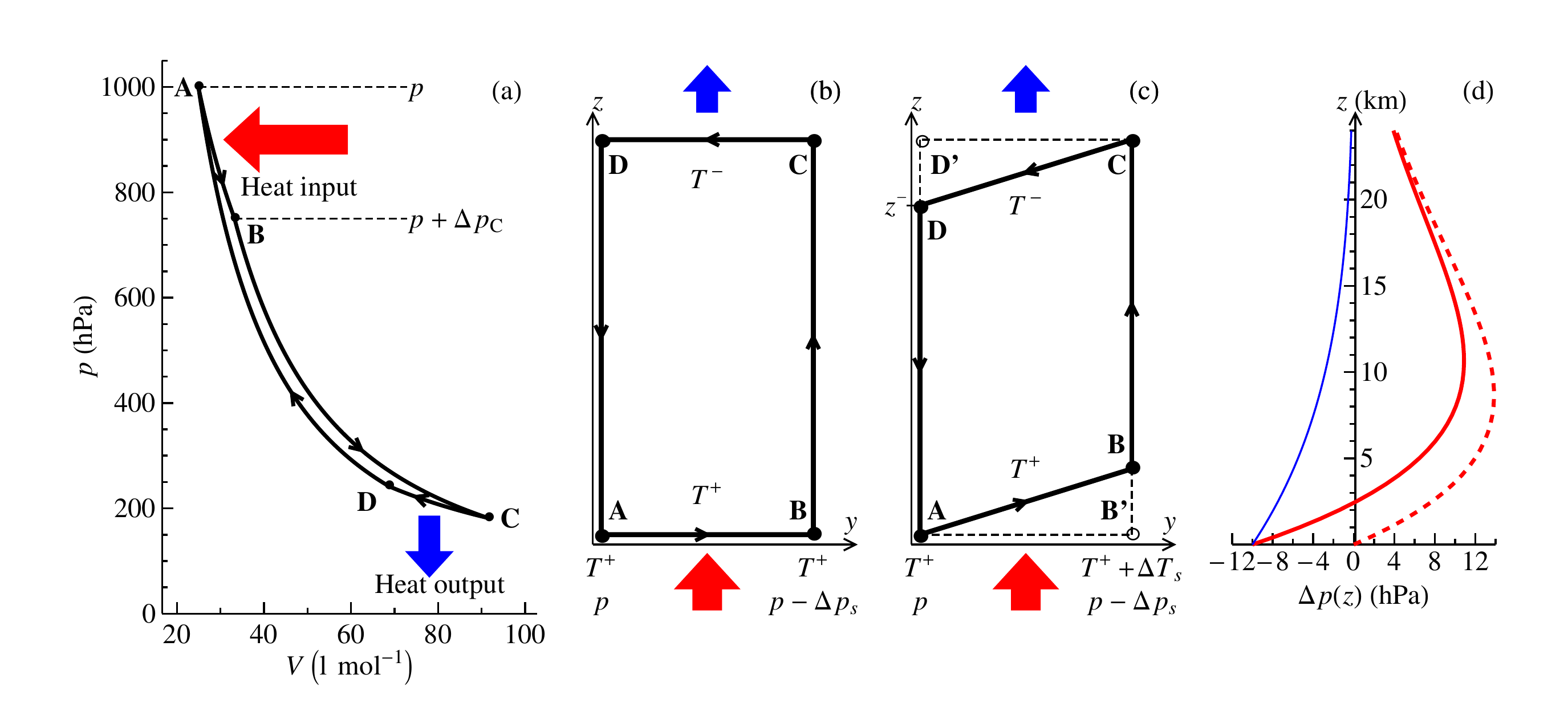}
}
\caption{\label{car}
Hydrostatic Carnot heat engine. (a) Carnot cycle with typical atmospheric parameters: $p = 1000$~hPa, $T^+ = 300$~K (isotherm $\rm AB$),
$T^- = 200$~K (isotherm $\rm CD$), $\Delta p_C = -250$~hPa. Work output at the warmer (colder) isotherm is equal to heat input (output).
BC and DA are adiabates. (b,c) The same cycle in spatial coordinates $y$, $z$ in (b) horizontally isothermal atmosphere ($\Delta T_s = 0$~K)
and (c) in the presence of a horizontal temperature difference ($\Delta T_s > 0$~K).
(d) Differences in pressure at height $z$ for atmospheric columns above points $\rm A$ and $\rm B'$, where pressure and temperature follow
Eqs.~(\ref{p}) and (\ref{T}) with surface pressure and temperature equal to $p_{\rm A} = 1010$~hPa, $T_{\rm A} = 300$~K
and $p_{\rm B'} = p_{\rm A} - \Delta p_s$, $T_{\rm B'} = T_{\rm A}+\Delta T_s$ at points $\rm A$ and $\rm B'$, respectively;
solid blue line: $\Delta p_s = 10$~hPa, $\Delta T_s = 0$~K; dashed red line:
$\Delta p_s = 0$~hPa, $\Delta T_s = 10$~K; solid red line: $\Delta p_s = 10$~hPa, $\Delta T_s = 10$~K.
}
\end{figure*}

We will now derive the dependence of kinetic energy generation on surface pressure and temperature differences in a Carnot cycle.
A Carnot cycle derives work from a given temperature gradient (heat engine) or uses external work to generate a temperature gradient (a heat pump). In theory the Carnot cycle requires that these processes are reversible and thus performed with maximum efficiency. In the atmosphere this  idealized cycle involves a heat flow from a warmer region (the heat source) to a colder region (the heat sink) in a
heat engine or in reverse direction in a heat pump. A Carnot cycle consists of two isotherms and two adiabates (Fig.~\ref{car}a). Work outputs on the two adiabates have different signs and sum to zero.
Total work output $A \equiv \oint pdV$, where $p$ is pressure and $V$ is molar volume, is equal to the sum of work outputs on the two isotherms. It can be written as \citep[for a derivation see, e.g.,][]{dhe10}
\begin{align}
A &= -R(T^+ - T^-) \ln \left(1+\frac{\Delta p_C}{p}\right) \approx -R(T^+ - T^-)\frac{\Delta p_C}{p}.  \label{Wtot}
\end{align}
Here $T^+$ and $T^- < T^+$ are the temperatures of the two isotherms, $p$ is the pressure that the air has as
it starts moving along the warmer isotherm and $\Delta p_C$ is the pressure change along the warmer isotherm, $|\Delta p_C|\ll p$. If the air
expands at the warmer isotherm (cycle $\rm ABCDA$), we have $\Delta p_C < 0$ and $A > 0$. The Carnot cycle is then a true heat engine: it
does work on the external environment and transports heat from the heat source to the heat sink (Fig.~\ref{car}a). If the air contracts at the warmer
isotherm (cycle BADCB) we have $\Delta p_C > 0$ and $A < 0$. The Carnot cycle functions now as a
heat pump: it transports heat from the heat sink (cold area) to the heat source (warm area) consuming work from the external environment.

We consider a closed streamline where the adiabates are vertical.
If the atmosphere is horizontally isothermal, then the isothermal parts of the streamline lie parallel to the surface
and the work output of the cycle is determined by surface pressure difference $\Delta p_C = -\Delta p_s$ (Fig.~\ref{car}b).
If there is a horizontal temperature gradient at the surface, then the isotherms are no longer horizontal,
but have an inclination that depends on the magnitude of the vertical temperature lapse rate
(Fig.~\ref{car}c). In this case $\Delta p_C$ depends on differences in surface pressure and temperature as well as on the lapse rate.

We consider a hydrostatic atmosphere with a constant lapse rate $\Gamma \equiv -\partial T/\partial z$, where
pressure and temperature depend on height $z$ and one horizontal coordinate $y$ \citep{jcli15}:
\begin{gather}
p(y,z) = p_s(y)\left[\frac{T(y,z)}{T_s(y)}\right]^{1/c}\equiv p_s(y)\left[1-\frac{\Gamma z}{T_s(y)}\right]^{1/c},  \label{p}\\
T(y,z) = T_s(y)-\Gamma z,\quad c \equiv \frac{\Gamma}{\Gamma_g},\quad  \Gamma_g \equiv \frac{Mg}{R}.        \label{T}
\end{gather}
Subscript $s$ denotes values of pressure and temperature on the geopotential surface;
$y$ represents distance along the meridian ($y \equiv a \varphi$, $a=1$ is the Earth's radius, $\varphi$ is latitude).
When the air moves from $y_1$ to $y_2$ following a streamline $z = z(y)$,
kinetic energy generation, as follows from Eqs.~(\ref{ig}) and (\ref{Wtotg}), is given by
\begin{equation} \label{Wz}
A = - RT \int_{y_1}^{y_2} \frac{1}{p(y,z)} \left.\frac{\partial p(y,z)}{\partial y}\right\vert_{z = z(y)} dy.
\end{equation}

Consider a cycle with positive total work as in Fig.~\ref{car}c.
Given the small relative changes of surface pressure and temperature we can write
their dependence on $y$ in the linear form:
\begin{equation}\label{psTs}
p_s(y) = p - \Delta p_s \frac{y}{\Delta y}, \quad  T_s(y) = T + \Delta T_s \frac{y}{\Delta y},
\end{equation}
where $p \equiv p_{\rm A}$, $T^+ \equiv T_{\rm A}$, $y_{\rm A} = 0$, $y_{\rm B'} = y_{\rm B} = \Delta y$, $\Delta p_s > 0$, $\Delta T_s > 0$.

Streamline equations for the warmer isotherm AB with temperature $T^+$ and the colder
isotherm CD with temperature $T^-$ are obtained from Eqs.~(\ref{psTs}) and (\ref{T}):
\begin{align} \label{z}
{\rm AB}:& \quad  z(y) = \frac{\Delta T_s}{\Gamma}\frac{y}{\Delta y}; \\
\label{z*}
{\rm CD}:&\quad z(y) = z^-+\frac{\Delta T_s}{\Gamma}\frac{y}{\Delta y},\quad  z^-\equiv \frac{T^+ - T^-}{\Gamma}.
\end{align}

Differentiating Eq.~(\ref{p}) over $y$ for constant $z$ and using Eqs.~(\ref{psTs})-(\ref{z*})
we obtain from Eq.~(\ref{Wz}) the following expressions for kinetic energy output $A^+$
and $A^-$ at the warmer and colder isotherms, respectively:
\begin{align}
\label{W}
A^+ &= RT^+\left[ \int_0^{\Delta p_s/p} \frac{dx}{1-x}\right. -\left. \frac{1}{c}\int_0^{\Delta T_s/T^+} \frac{xdx}{1+x}\right],\\
\label{W*}
A^- &= RT^-\left[ -\frac{A}{RT^+} + \frac{1}{c}\int_0^{\Delta T_s/T^+} \left(\frac{T^+}{T^-}-1\right) \frac{dx}{1+x}\right].
\end{align}

As we will discuss in greater detail below, relative differences $\Delta p_s/p_s$ and $\Delta T_s/T_s$ on Earth possess small magnitudes
of the order of $\Delta p_s/p_s \sim 10^{-2}$  and $(1/c)(\Delta T_s/T_s)^2 \sim 10^{-2}$. Therefore,
all calculations in Eqs.~(\ref{W}) and (\ref{W*}) can be done to the accuracy of the linear terms over
$\Delta p_s/p$ and linear and quadratic terms over $\Delta T_s/T^+$.
Performing integration in Eqs.~(\ref{W}) and (\ref{W*}) and expressing the result in these approximations we obtain:
\begin{align}
A^+ &= RT^+ \left[\frac{\Delta p_s}{p} - \frac{1}{2c} \left(\frac{\Delta T_{s}}{T^+}\right)^2\right] , \,\,\,
A_C^+ = A^+ + \frac{R\Delta T_s}{c}, \label{WK}\\
A^- &= -\frac{T^-}{T^+} A^+ +\frac{R(T^+-T^-)}{c} \frac{\Delta T_s}{T^+}, \,\,\,
A_C^-= -\frac{T^-}{T^+}\left(A^+ + \frac{R\Delta T_s}{c}\right),  \label{WK*}
\end{align}
\begin{align}
A &= -R(T^+-T^-) \ln \left[ \frac{1 - \Delta p_s/p}{(1+\Delta T_s/T^+)^{1/c}} \right]  = A_C^+ + A_C^- = A^+ + A^-.  \label{wtot}
\end{align}
Here $A_C^+ \equiv \int_{V_{\rm A}}^{V_{\rm B}} pdV$ and $A_C^- \equiv \int_{V_{\rm C}}^{V_{\rm D}} pdV$ are the total work outputs on isotherms
AB and CD, respectively (italic subscript "$C$" stands for "Carnot"); $p \equiv p_{\rm A}$ and $T^+ \equiv T_{\rm A}$ are surface pressure and temperature at point $\rm A$,
$p - \Delta p_s \equiv p_{\rm B'}$ and $T^+ +\Delta T_s \equiv T_{\rm B'}$ are pressure and temperature at point
$\rm B'$. Eq.~(\ref{wtot}) can also be obtained from Eq.~(\ref{Wtot}) with use
of Eqs.~(\ref{p}) and (\ref{T}) and noting that $\Delta p_C = p_{\rm B} - p_{\rm A}$ and $z_{\rm B} = \Delta T_s/\Gamma$.

Let us discuss the physical meaning of the obtained expressions.
We first note that kinetic energy generation $A^+$, $A^-$ and total work $A_C^+$, $A_C^-$
on the two isotherms may have different signs. In particular, at the warmer isotherm kinetic energy generation $A^+$ can be
either positive (at small $\Delta T_s$) or negative (at large $\Delta T_s$), while total work
$A_C^+$ is always positive at large $\Delta T_s>0$, which reflects the fact that at the lower isotherm the gas expands.

For horizontally isothermal surface with $\Delta T_s = 0$ (Fig. \ref{car}b) we have $A^+> 0$ and $A^-<0$,
i.e. kinetic energy generation at the colder isotherm is negative.
It means that in the upper atmosphere the air must move
in the direction of growing pressure (see Fig. \ref{car}d, solid blue line) thus losing kinetic energy.
At the beginning of this path (at point $\rm C$) the air must possess sufficient kinetic energy exceeding $|A^-| = RT^- \Delta p_s/p$
to cover the entire isotherm $\rm CD$.
If the initial store of kinetic energy is insufficient, at a certain point between $\rm C$ and $\rm D$
the kinetic energy becomes zero. Not having reached point $\rm D$,
the air will start moving in the opposite direction under the action of the pressure gradient force.
For $T^- = 200$~K and $\Delta p_s = 10$~hPa we have $A^- = -17$~J~mol$^{-1}$ or $A^-/M = -570$~J~kg$^{-1}$,
where $M = 29$~g~mol$^{-1}$ is molar mass of air.
This corresponds to an air velocity of about $\sqrt{|A^-|/M} \sim 24$~m~s$^{-1}$,
which is a typical velocity in the upper troposphere. Thus the cycle
shown in Fig. \ref{car}b is energetically plausible.

Now consider a situation when $\Delta p_s = 0$, but $\Delta T_s > 0$.
The relationship between work outputs is reversed: kinetic energy generation is negative at
the warmer isotherm, $A^+ < 0$, and positive at the colder isotherm, $A^->0$. This is because
at all heights, except at the surface, air pressure is higher
in the warmer area towards which the low-level air is moving (Fig.~\ref{car}d, red dashed line).
Thus, as before at the colder isotherm,
now the air must spend its kinetic energy to overcome the opposing action of the horizontal pressure gradient
force at the warmer isotherm. Using a typical value of $\Delta T_s \sim 10$~K in the Hadley cells for $T^+ = 300$~K,
$\Gamma = 6$~K~km$^{-1}$ ($2c = 0.35$, see~Eq.\eqref{T})
from Eq.~(\ref{WK}) we obtain $A^+/M = -270$~J~kg$^{-1}$. This means that the necessary velocity the air must possess at the beginning of the warmer isotherm to be able to cover it all from point $\rm A$ to point $\rm B$ is $\sqrt{|A^-|/M} \sim~16$~m~s$^{-1}$. This exceeds the characteristic
velocities at circulation cell boundaries in the boundary layer, which are about 8~m~s$^{-1}$ \citep[e.g.,][Fig.~1a]{lindzen87,schneider06}. Since in the lower atmosphere the air must also overcome surface friction, total energy required to move from $\rm A$ to $\rm B$ with $\Delta p_s = 0$
is larger than estimated from Eq.~(\ref{WK}).

If the kinetic energy the air possesses is negligible
compared to what is needed to move from $\rm A$ to $\rm B$,
the kinetic energy required must be generated on the warmer isotherm.
For kinetic energy generation on the warmer isotherm to be positive, $A^+ >0$,
surface pressure and temperature differences $\Delta p_s$ and $\Delta T_s$ must satisfy
\begin{equation}
\label{cond}
K \equiv 2c\frac{\Delta p_s}{p} \left(\frac{T^+}{\Delta T_s}\right)^2 > 1.
\end{equation}

\begin{figure}
\centering\includegraphics[width=0.48\textwidth,angle=0,clip]{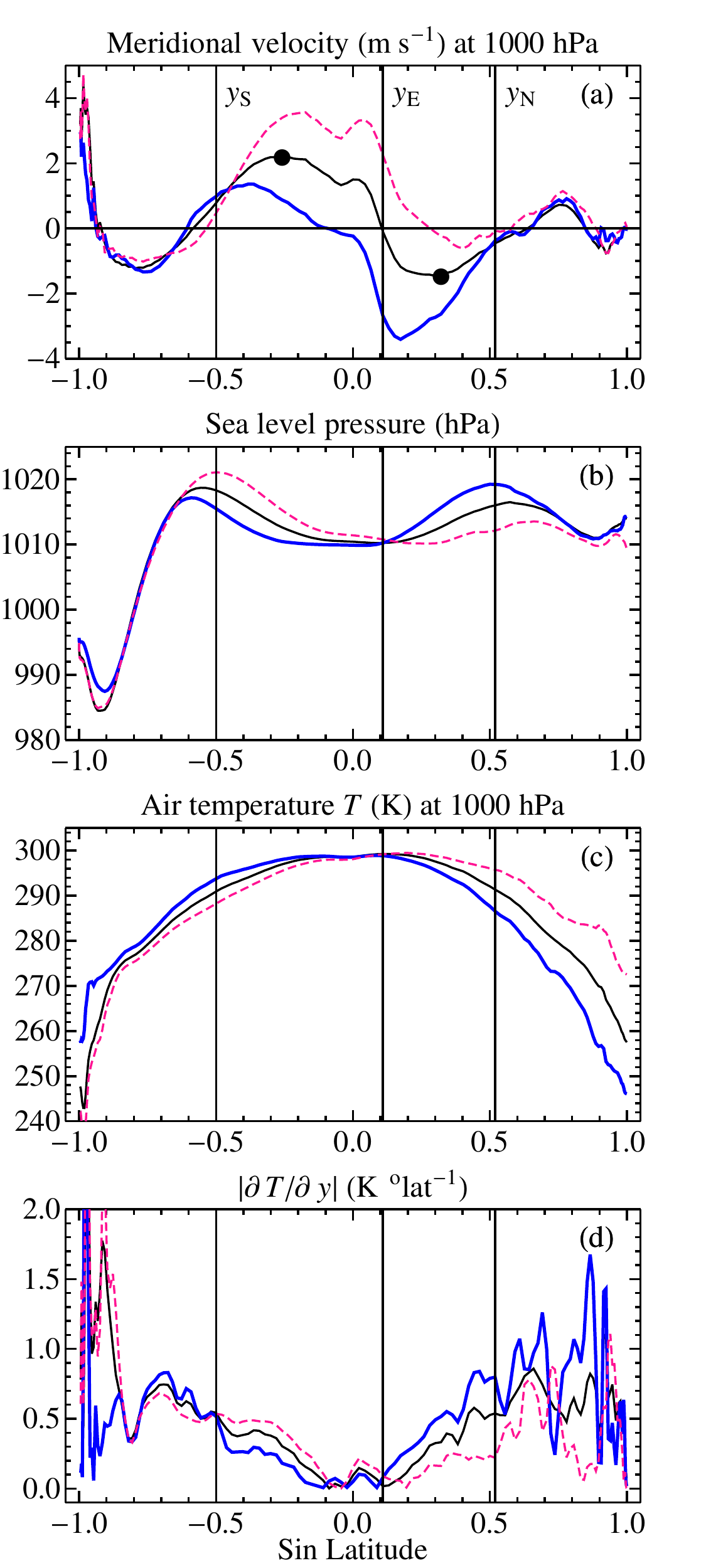}
\caption{\label{had}
Long-term mean zonally averaged meridional velocity, sea level pressure (SLP), air temperature $T$ at 1000 hPa and
its meridional gradient calculated from MERRA data. Black, thick blue and pink dashed curves denote
annual mean, January and July data, respectively. Black circles in (a) indicate
velocity maxima; $y_{\rm N}$, $y_{\rm S}$ and $y_{\rm E}$ are the Hadley cell outer and inner borders that are calculated for each of
the 432 months from the velocity data (see text for details), here exemplified for the annual mean velocity curve.
For each month $\Delta p_{sj} \equiv {\rm SLP}(y_j) - {\rm SLP}(y_{\rm E})$, $\Delta T_{sj} \equiv T(y_{\rm E}) - T(y_j)$, where $j=\rm S,N$ stand for the Southern
and Northern cells,
respectively.
}
\end{figure}

For $T^+ = 250$~K and $\Delta T_s = 50$~K, which characterize the surface temperature
differences between the equator and the poles, at $p = 1000$~hPa
and $\Gamma = 6.5$~K~km$^{-1}$ ($2c = 0.38$) we find that to satisfy Eq.~(\ref{cond})
the surface pressure difference between the equator and the pole
$\Delta p_s$ must exceed 100~hPa. Such a pressure difference on Earth can be found in intense
compact vortices like the severest hurricanes and tornadoes; it is an order
of magnitude larger than the typical $\Delta p_s$ in the two Hadley cells (Fig.~\ref{had}b),
which together cover over half of the Earth's surface.

From Eq.~(\ref{cond}) we can derive the maximum size of the cell expressed in terms
of the absolute magnitudes of surface pressure and temperature gradients $\nabla p_s \equiv \partial p_s/\partial y
=-\Delta p_s/\Delta y$,
$\nabla T_s \equiv \partial T_s/\partial y =\Delta T_s/\Delta y$:
\begin{equation}
\label{max}
\Delta y \le L_{\mathrm{max}} \equiv 2c \frac{|\nabla p_s|}{p} \left(\frac{T^+}{\nabla T_s}\right)^2.
\end{equation}

For the Carnot cycle, the maximum cell size $L_{\mathrm{max}}$ grows with increasing surface pressure gradient and diminishes
proportionally to the squared surface temperature gradient. The surface temperature
gradient is largely known: it reflects the differential solar heating that makes the pole colder
than the equator. Therefore, the key question in deciding about the maximum cell size is what
determines the surface pressure gradient.

\section{Thermodynamic cycle with rectangular streamlines}

In the Carnot cycle, the $\Delta T_s$ term in the expression for kinetic energy generation $A^+$ is quadratic,
see Eqs.~(\ref{WK}),  (\ref{cond}), (\ref{max}). This results from the dual nature of the relationship.
Firstly, $\Delta T_s$ determines the mean height of the warmer isotherm via the streamline equation Eq.~(\ref{z}):
the greater $\Delta T_s$, the greater the mean height at which the air moves in the lower atmosphere.
Secondly, $\Delta T_s$ determines a pressure difference that acts as a sink for kinetic energy in the lower
atmosphere (see Fig.~\ref{car}d, dashed line). This pressure difference increases linearly with small $\Delta T_s$ and with height.
This double effect leads to the negative quadratic term in Eq.~(\ref{WK}), which diminishes the rate at which kinetic energy
is generated in a heat engine.

We will now consider kinetic energy generation $A_r$ in a cycle where the vertical adiabates are connected by horizontal streamlines. The air moves  parallel to the surface at a certain height $z$ that is independent of $\Delta T_s$. It is a cycle with streamlines
as shown in Fig.~\ref{car}b but in the presence of a surface temperature difference as in Fig.~\ref{car}c.

Using Eqs.~(\ref{p}), (\ref{T}) and (\ref{psTs}) we find that
in this case kinetic energy generation at the horizontal streamlines is given by
\begin{align}
A_r(y_1,y_2,z) =&  -\int_{p_1}^{p_2} \frac{RT}{p}dp  = -R\int_{y_1}^{y_2} \frac{T_s(y)-\Gamma z}{p} \frac{\partial p}{\partial y}dy \label{Wr}\\
=& -R\Gamma z \int_{y_1}^{y_2}  \left(-\frac{1}{p_s}\frac{\partial p_s}{\partial y}  +
  \frac{1}{c T_s}\frac{\partial T_s}{\partial y} \right)dy   -R\int_{y_1}^{y_2} \frac{T_s}{p_s} \frac{\partial p_s}{\partial y} dy \\
= &  R\Gamma z \ln \left[ \frac{1 - \Delta p_s/p}{(1+\Delta T_s/T^+)^{1/c}} \right]   - RT^+ \ln \left(1-\frac{\Delta p_s}{p}\right)  -R\Delta T_s\left[ 1+\frac{p}{\Delta p_s} \ln \left(1-\frac{\Delta p_s}{p}\right)\right] .  \label{Wr2}
\end{align}
At the lower AB and upper CD streamlines, kinetic energy generation is calculated from Eq.~(\ref{Wr2}) as follows:
$A_r^+ \equiv A_r(y_{\rm A},y_{\rm B},z^+)$ and $A_r^- \equiv - A_r(y_{\rm C},y_{\rm D},z^-)$, respectively, where $z^+$ and $z^- > z^+$
are the altitudes of these streamlines, $y_{\rm C} = y_{\rm B}$, and $y_{\rm D} = y_{\rm A}$.

Retaining linear terms over $\Delta p_s/p$ and $\Delta T_s/T^+$ (thus discarding the last term in Eq.~\eqref{Wr2})
for $A_r^+$ and $A_r^-$ we find
\begin{align}
A_r^+ =& R(T-\Gamma z^+) \frac{\Delta p_s}{p} -R\Gamma z \frac{1}{c}\frac{\Delta T_s}{T^+}  = RT \frac{\Delta p_s}{p} -M g z^+ \frac{\Delta T_s}{T^+},   \label{Wrr}\\
A_r^- =& -R(T-\Gamma z^-)\frac{\Delta p_s}{p} + Mg z^- \frac{\Delta T_s}{T^+},  \label{Wrr*} \\
A_{r} \equiv & A_r^+ + A_r^-  = -R\Gamma (z^- - z^+) \ln \left[ \frac{1 - \Delta p_s/p}{(1+\Delta T_s/T^+)^{1/c}} \right].  \label{Wrtot}
\end{align}
We additionally took into account that $\Gamma z^+ \ll T^+$ in the second equality in Eq.~(\ref{Wrr}). Note that
$R\Gamma/c = Mg$ (see Eq.~\eqref{T}).

With $z^+ = \Delta T_s/(2\Gamma)$ equal to the mean height of an isothermal streamline in a Carnot cycle,
Fig.~\ref{car}c, Eq.~(\ref{Wrr}) coincides with Eq.~(\ref{WK}): $A^+ = A_r^+$.
For the upper streamline Eq.~(\ref{Wrr*}) coincides with $A^-$, Eq.~(\ref{WK*}),
if we put $z^- - z^+ \approx z^- = (T^+ - T^-)/\Gamma$, where $T^+ \equiv T_{\rm A}$ and $T^- \equiv T_{\rm D}$, Fig.~\ref{car}.
Under this assumption total work $A_r$ of this cycle coincides with total work of the Carnot cyle, Eq.~(\ref{wtot}).
(It can be shown using Eqs.~(\ref{p}) and (\ref{T}) that the efficiency of the "rectangular" cycle is lower by a small magnitude of
the order of $\Gamma z^+/T$.)
As in the Carnot cycle, the contribution of temperature difference $\Delta T_s$ to kinetic energy generation is negative
at the lower streamline, cf. Eqs.~(\ref{Wrr}) and (\ref{WK}), and positive at the upper streamline, cf. Eq.~(\ref{Wrr*}) and (\ref{WK*}).
But the negative contribution is now linear over $\Delta T_s$, not quadratic.

From Eq.~(\ref{Wrr}) the condition that $A_r^+ > 0$ takes the form
\begin{align}\label{Wrze}
&A_r^+ = Mgh_s \left(\frac{\Delta p_s}{p}  - \frac{z}{h_s} \frac{\Delta T_s}{T^+}\right) > 0,\quad
h_s \equiv \frac{RT^+}{Mg},\\
&z^+ < z_e \equiv h_s \frac{\Delta p_s}{p}\frac{T^+}{\Delta T_s}. \label{ze}
\end{align}
Here $h_s$ is the atmospheric scale height and $z_e$ is the isobaric height at which the pressure difference between two
atmospheric columns turns to zero \citep{jcli15}.  Equation~(\ref{ze}) indicates that the height where the low-level air moves must be smaller than the isobaric height. Since $\Delta T_s$ increases as the cell extends towards the pole, for $z^+$ to be constant, $\Delta p_s$ must grow approximately
proportionally to $\Delta T_s$. If $\Delta p_s$ grows more slowly than $\Delta T_s$, then at a certain $\Delta T_s$ height
$z^+$ can turn to zero or, at constant $z^+$, $A_r^+$ becomes negative. The dependence between $\Delta p_s$ and $\Delta T_s$ thus dictates the maximum
cell size as long as $A_r^+$ is positive.  In Appendices~B and C we discuss how Eq.~(\ref{Wrze}) and its implications are impacted by our
assumptions of idealized air trajectories and constant lapse rate.

\section{Surface pressure, temperature and cell size in Hadley cells}

To investigate the dependence between $\Delta p_s$ and $\Delta T_s$ in Hadley cells
we used monthly averaged sea level pressure (SLP) and meridional velocity at 1000 hPa from
1979 to 2014 from MERRA re-analysis (a total of 432 months), which has a resolution of $1.25\degree\times 1.25\degree$ ($144 \times 288$ grid cells).
For each month we established
the long-term mean position of the maxima of zonally averaged SLP at
the outer borders of the cells. Then for each month in each year
we defined the inner border of the cells $y_{\rm E}$ as the latitude
of minimum zonally averaged SLP located between the long-term maxima
(Fig.~\ref{had}b). SLP values were considered different if they differed by not less than 0.05 hPa.
If there were several minimal pressure values equal to each other, we chose the one where the
absolute magnitude of the zonally averaged meridional
velocity was minimal.

Then for each cell for each month we calculated maximum (by absolute magnitude) velocity
poleward from the inner border (Fig.~\ref{had}a). The outer border of the Northern (N) and
Southern (S) cell $y_{\rm N}$ and $y_{\rm S}$ was
defined as the latitude poleward of maximum meridional velocity where
the velocity declined by $e$ times (or changed sign) as compared to the maximum. This relative threshold
was chosen because the meridional velocity distributions within the two cells are
different: in the Northern hemisphere the seasonal change of meridional velocities
is greater than it is in the Southern hemisphere (Fig.~\ref{had}a). As we are
interested in kinetic energy  generation that depends on velocity, we need to define the cell borders relevant
to velocity. Temperature changes in the regions where velocity is close
to zero do not impact the kinetic energy generation.

\begin{figure*}
\centerline{
\includegraphics[width=0.99\textwidth,angle=0,clip]{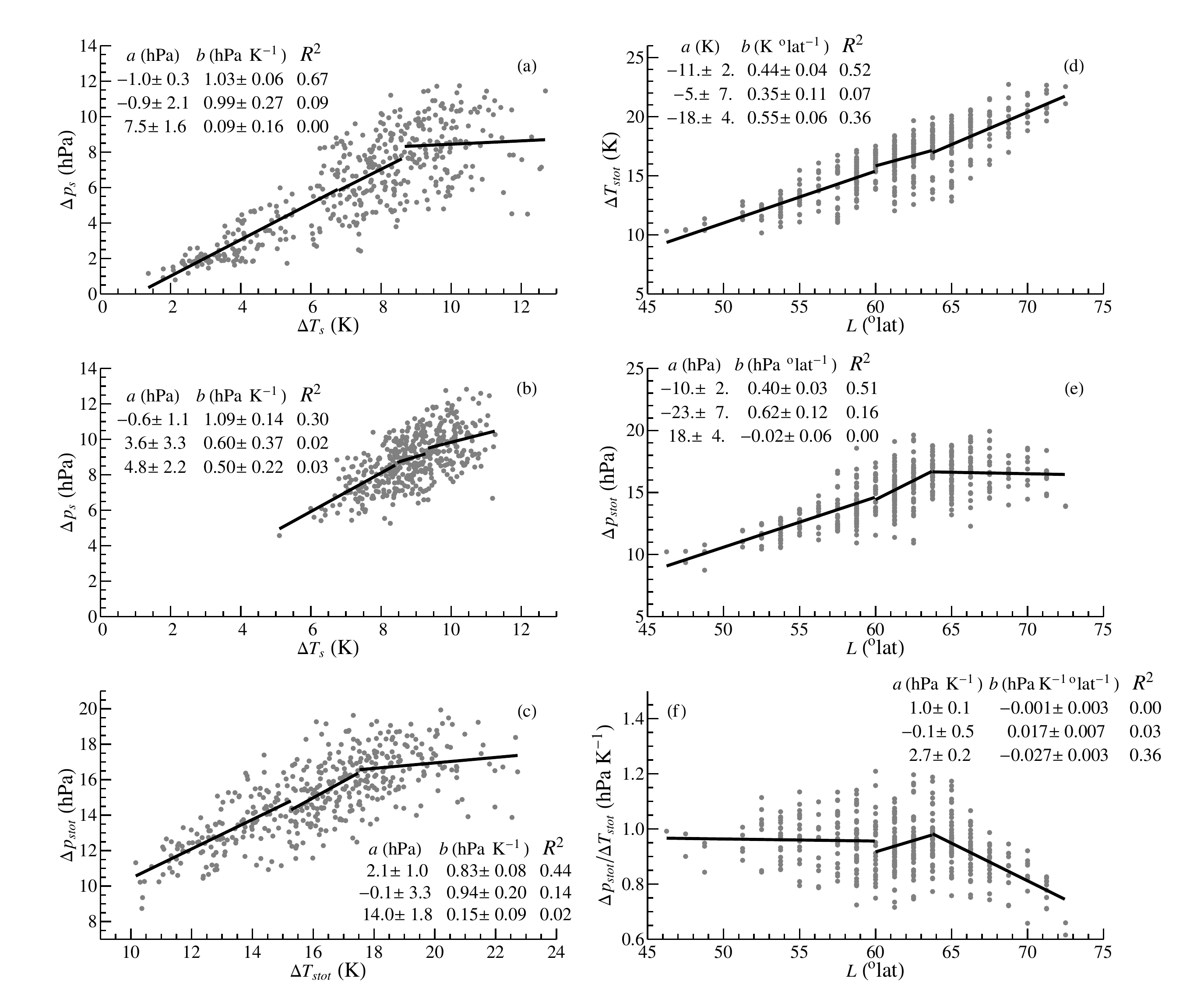}
}
\caption{\label{figdpdt}
Relationships between surface temperature and pressure differences and the meridional extension of the Hadley system.
(a-c): Dependence of $\Delta p_s$ on $\Delta T_s$ in (a) Northern Hadley cell, (b) Southern Hadley cell,
(c) Hadley system as a whole:
$\Delta p_{s\mathrm{tot}} \equiv \Delta p_{s\rm N} + \Delta p_{s\rm S} = p_s(y_{\rm N})+p_s(y_{\rm S})-2p_s(y_{\rm E})$,
$\Delta T_{s\mathrm{tot}} \equiv \Delta T_{s\rm N} + \Delta T_{s\rm S} = T_s(y_{\rm N})+T_s(y_{\rm S})-2T_s(y_{\rm E})$
(Fig.~\ref{had}); (d-f): dependence of $\Delta T_{s\mathrm{tot}}$ (d), $\Delta p_{s\mathrm{tot}}$ (e)
and their ratio (f) on the total extension of the Hadley system ($L \equiv y_{\rm N} + |y_{\rm S}|$) (degrees latitude).
Solid lines denote linear regressions $y = a x + b$ for the 144 lowest, 144 intermediate and 144 highest
values of $x=\Delta T_s$ (a,b), $x=\Delta T_{s\mathrm{tot}}$ (c) and $x=L$ (d-f). Regression
parameters are shown in each panel starting from the lowest $x$ values.
}
\end{figure*}

We find that both in the Northern and Southern cells the dependence between $\Delta p_s$ and $\Delta T_s$ changes
with growing $\Delta T_s$. For the Northern cell for the 144 lowest,144 intermediate and 144 highest values of $\Delta T_s$ the linear
regression $\Delta p_s = a + b \Delta T_s$ yielded, respectively, $b = 1.03$, $0.99$ and $0.09$~hPa~K$^{-1}$; for the Southern cells it was
$1.09$, $0.60$ and $0.50$~hPa~K$^{-1}$ (Fig.~\ref{figdpdt}a,b).
While for the smaller values of $\Delta T_s$ the relationship in both cells is identical and the
proportionality coefficient is about $1$~hPa~K$^{-1}$, it decreases markedly (in the Northern cell -- down
to zero) with growing $\Delta T_s$. At constant $z^+$ this results in declining kinetic energy generation (Eq.~\eqref{Wrr}).

\begin{figure*}
\centerline{
\includegraphics[width=0.6\textwidth,angle=0,clip]{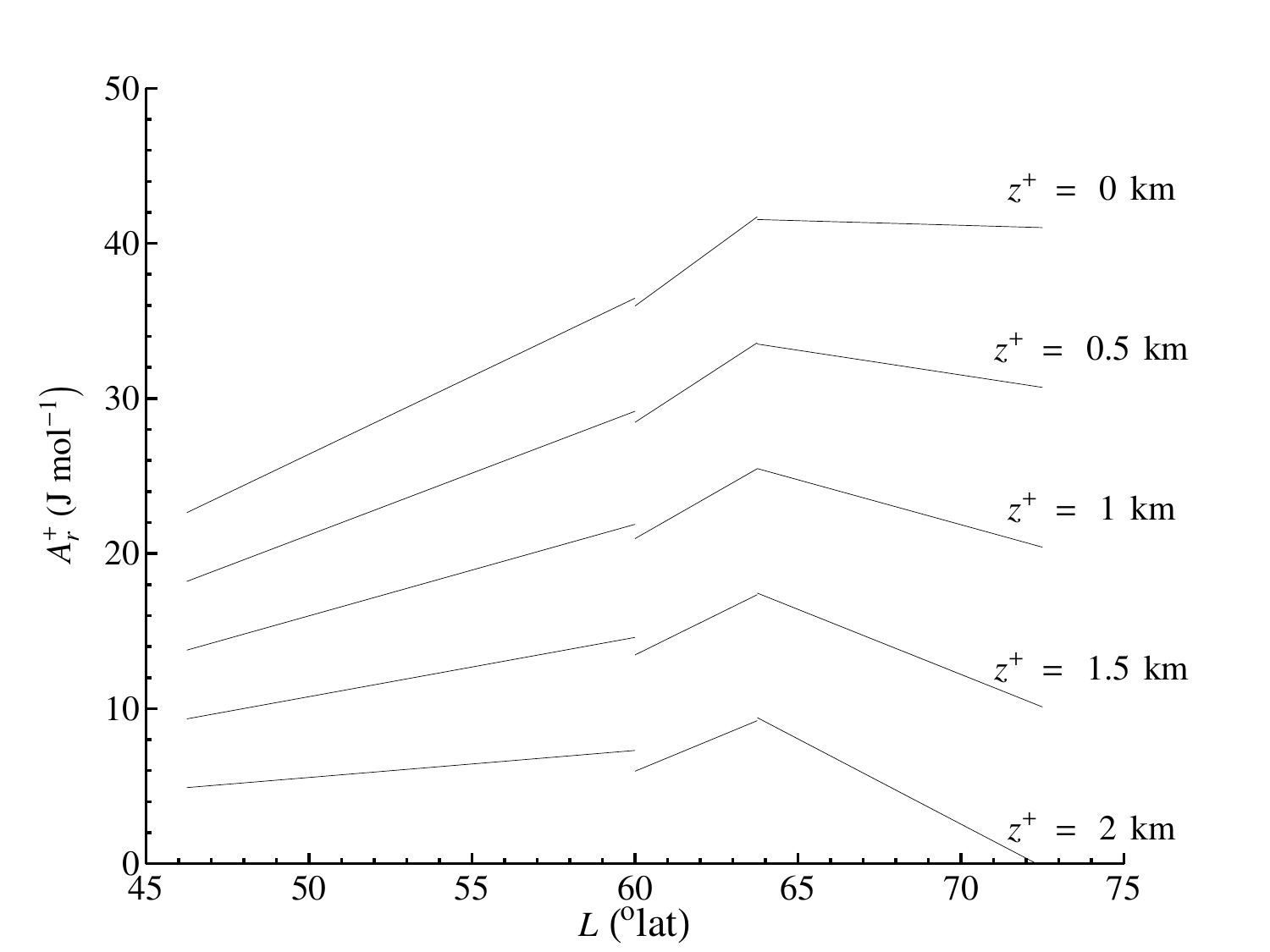}
}
\caption{\label{figAr}
Kinetic energy generation $A_r^+$ at different altitudes $z^+$ in the lower atmosphere
versus Hadley system size $L$ as determined from Eq.~(\ref{Wrr}) using the observed
$\Delta p_{s\mathrm{tot}}$ and $\Delta T_{s\mathrm{tot}}$ values from Fig.~\ref{figdpdt}d,e.
}
\end{figure*}

\begin{figure*}
\begin{minipage}[h]{0.37\textwidth}
\centerline{
\includegraphics[width=0.99\textwidth,angle=0,clip]{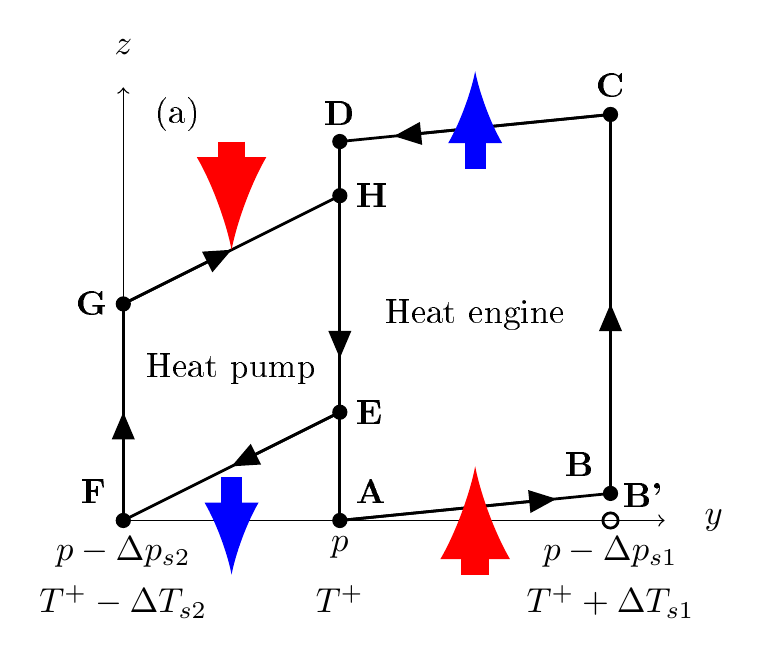}
}
\end{minipage}
\begin{minipage}[h]{0.6\textwidth}
\centerline{
\includegraphics[width=0.99\textwidth,angle=0,clip]{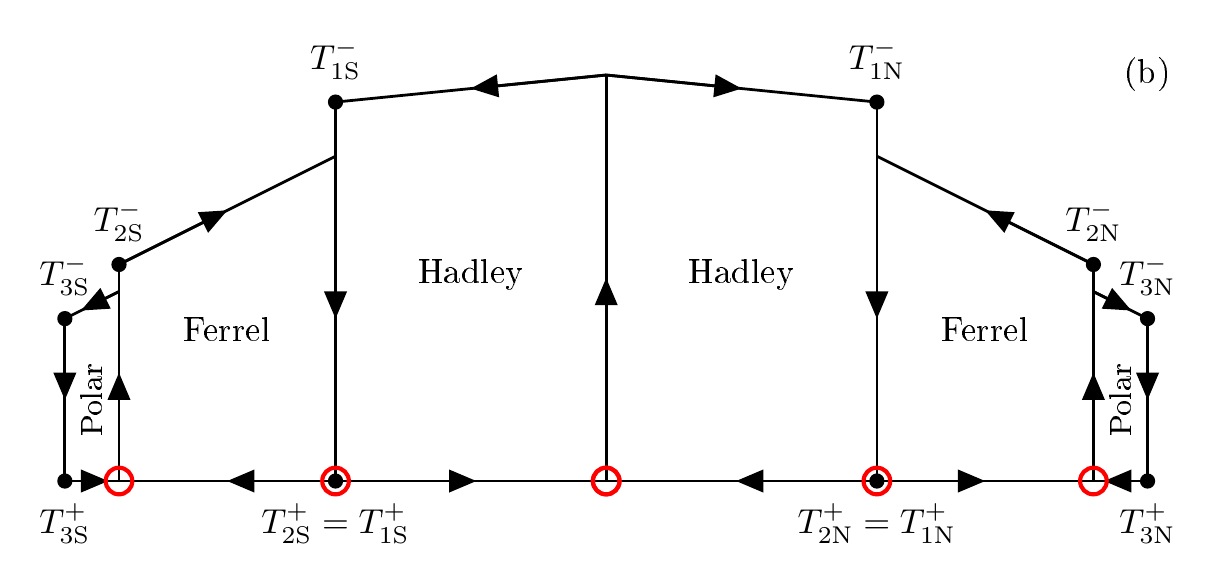}
}
\end{minipage}
\caption{\label{cells}
Heat pumps and heat engines. (a) A Carnot heat pump (cycle $\rm EFGHE$) bordering with a Carnot heat engine (cycle $\rm ABCDA$). See Fig.~\ref{car} for
details. (b) The meridional circulation cells on Earth. Arrows show the direction of air movement.
Empty red circles indicate the latitude of cell borders as determined in Fig.~\ref{figfri}a from meridional velocity values.
Black circles indicate where temperature and pressure values used in Eqs.~(\ref{Wg}) and (\ref{Wg*}) were calculated.
The vertical dimension is height $z$, the horizontal dimension is sine latitude which accounts for cell area.
}
\end{figure*}

We further observe that while temperature difference $\Delta T_s$ grows with the cumulative extension $L \equiv y_{\rm N} - y_{\rm S}$
of the Hadley system, the surface pressure difference reaches a plateau of 17 hPa for $L \approx 64\degree$~latitude (Fig.~\ref{figdpdt}d,e).
Accordingly, for $L < 64\degree$
the ratio of cumulative pressure and temperature differences for the Hadley system
is essentially constant at around $1$~hPa~K$^{-1}$, but for larger $L$ it
declines by about one third as $L$ grows up to the maximum observed values (Fig.~\ref{figdpdt}f).
Figure~\ref{figAr} describes kinetic energy generation $A_r^+$ at different altitudes in the boundary layer.
For $z^+ > 0$ kinetic energy generation declines for $L>64\degree$~latitude. For $z^+ = 2$~km $A_r^+$ becomes
zero at the observed maximum extension of the Hadley system $L_{\mathrm{max}} = 72.5\degree$~latitude.

\section{Heat engines and heat pumps}
\label{hp}

If the poleward extension of a cell with positive total work output (a heat engine like the Hadley cell) is limited,
a cell with a negative work output (working as a heat pump) must be present poleward of the
heat engine (Fig.~\ref{cells}). This is a requirement of continuity: in the adjacent cells the air must go
downwards along the path DA. Hence, in the lower atmosphere the air must move from the warmer to the colder region.
Such a heat pump requires an external supply of work, which can be provided by
neighboring heat engines that have a positive work output. In the atmosphere of the Earth
Ferrel cells are the heat pumps, while the Hadley and Polar cells are the heat engines \citep[see, e.g.,][]{huang14}.

In the lower atmosphere the global kinetic energy generation for the six cells can be
written using Eq.~(\ref{Wrr}) as
\begin{equation}
\label{Wg}
A_g^+ = \sum_{j={\rm S,N}} \sum_{i=1}^3 a_{ij} \left[(-1)^{i} Mgz \frac{\Delta T_{ij}}{T_{ij}^+} + RT_{ij}^+ \frac{\Delta p_{ij}}{p_{ij}}\right].
\end{equation}
Here summation is over the six cells, three in the Southern ($j=\rm S$) and three in the Northern ($j=\rm N$) hemisphere,
$i = 1,2,3$ for Hadley, Ferrel and Polar cells, respectively; $\Delta T_{ij}\equiv \Delta T_{sij}>0$ and
$\Delta p_{ij}\equiv \Delta p_{sij}>0$ stand for the differences in surface temperature and pressure within the cell;
$T_{ij}$ and $p_{ij}$ are surface temperature and pressure at the beginning of the lower streamline (Fig.~\ref{cells}).
Note that $T_{1j}^+ = T_{2j}^+$ and $p_{1j}^+ = p_{2j}^+$ (the lower streamlines in Hadley and Ferrel
cells start from the same latitude). We have accounted for the area by introducing coefficients $a_{ij} = \sin y_{{\rm P}ij} - \sin y_{{\rm E}ij}$,
which is the relative area of the cell, where $y_{\rm P}$ and $y_{\rm E}$ are the latitudes of cell borders closest to the pole (P) and equator (E)
(for Polar cells $y_{\rm P} = \pm 90\degree$). Thus work output of each cell is weighted by area.

Coefficient $(-1)^{i}$ differentiates heat engines ($i=1,3$) from heat pumps ($i=2$) (Table~\ref{contrib}).
For a "rectangular" heat pump where low surface pressure is associated with low surface temperature, such that
the air compresses at the lower streamline, the temperature terms in Eqs.~(\ref{Wrr}) and (\ref{Wrr*}) change their sign.
(Note that the pressure terms do not change their sign if, as in the heat engine, the low level air moves from high
to low surface pressure.) In a heat pump the temperature difference makes a negative contribution to
kinetic energy generation in the upper atmosphere and a positive contribution in the lower atmosphere (Table~\ref{contrib}).
Thus kinetic energy generation in the lower atmosphere can be expected to be greatest in heat pumps, which agrees
with observations (see Table~\ref{number} below).

\begin{table*}
\scriptsize
\begin{minipage}[h]{0.99\textwidth}
\caption{\label{contrib}Contributions of surface pressure and temperature differences
to kinetic energy generation for hybrid cycles that have isothermal streamlines
in the upper atmosphere (Eq.~\eqref{WK*}) and horizontal streamlines in the lower
atmosphere (Eq.~\eqref{Wrr}). (For heat pumps the expression is valid
for $\Delta T_s \ll T^+- T^-$, cf. Eq.~\eqref{ferrel}).
}
\begin{center}
\begin{tabular}{lccc}
\toprule
&	Pressure &	\multicolumn{2}{c}{Temperature} \\
\cmidrule(r){2-2} \cmidrule(r){3-4}
&	All cells &  Heat engines	 & Heat pumps \\
\cmidrule(r){2-2}\cmidrule(r){3-3}\cmidrule(r){4-4}
&$ $ &&\\
Upper atmosphere &   $\displaystyle -RT^-\frac{\Delta p_s}{p} $  &	 $\displaystyle +\frac{R}{c}\left[(T^+-T^-)\frac{\Delta T_s}{T^+}+\frac{T^-}{2}\left(\frac{\Delta T_s}{T^+}\right)^2\right]$ &  $\displaystyle -\frac{R}{c}\left[(T^+-T^-)\frac{\Delta T_s}{T^+}+\frac{T^-}{2}\left(\frac{\Delta T_s}{T^+}\right)^2\right]$ \\
&$ $ &&\\
Lower atmosphere &   $\displaystyle +RT^+\frac{\Delta p_s}{p}    $  &	 $\displaystyle -Mgz^+\frac{\Delta T_s}{T^+}$ &	 $\displaystyle +Mgz^+\frac{\Delta T_s}{T^+}$ \\
&$ $ &&\\
\bottomrule
\end{tabular}
\end{center}
\end{minipage}
\end{table*}

To estimate global kinetic energy generation in the upper atmosphere we use Eq.~(\ref{WK*}) for the colder isotherm of the Carnot cycle (Fig.~\ref{cells}b):
\begin{align}
\label{Wg*}
A^-_g =& \sum_{j={\rm S,N}} \left[ \sum_{i=1,3}
a_{ij} A_{ij}^- + a_{2j} A_{2j}^-\right],\\
A_{2j}^- =& -RT_{2j}^-\left[\frac{\Delta p_{2j}}{p_{2j}}+\frac{1}{2c} \left(\frac{\Delta T_{2j}}{T_{2j}^+}\right)^2\right]
- R(T_{2j}^+ - \Delta T_{2j} - T_{2j}^-)\frac{1}{c}\frac{\Delta T_{2s}}{T_{2j}^+}.   \label{ferrel}
\end{align}
Here $A_{ij}^-$ for $i=1,3$ is work output of the heat engines (Polar and Hadley cells), Eq.~(\ref{WK*});
$A_{2j}^- <0$ is work output of the heat pumps (Ferrel cells)\footnote{Note that Eq.~(\ref{ferrel}) is not completely symmetrical to
Eq.~(\ref{WK*}), because in a heat pump temperature
of the warmer isotherm $T_{2j}^+$ is not equal to surface temperature $T_{2j}^+-\Delta T_{2j}$ at the beginning of the
warmer isotherm (which is point $\rm E$ in Fig.~\ref{cells}a). Meanwhile in the heat engine ($\rm ABCDA$) temperature of the warmer isotherm coincides
with surface temperature at point $\rm A$.}. It can be obtained following the procedure
that yielded Eq.~(\ref{WK*}) but for the isotherm $\rm GH$ instead of $\rm CD$
(Fig.~\ref{cells}a).

\begin{figure*}
\centerline{
\includegraphics[width=0.8\textwidth,angle=0,clip]{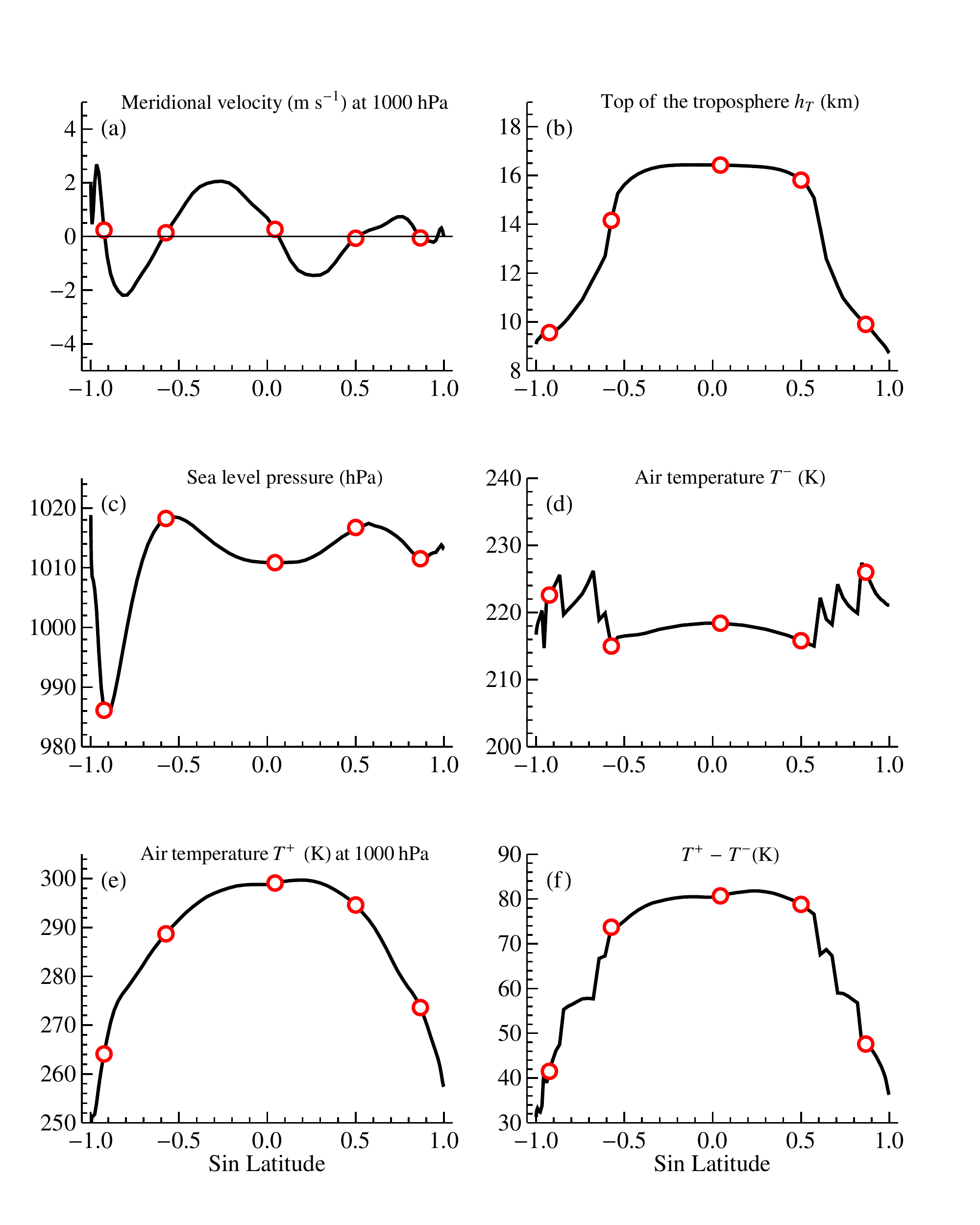}
}
\caption{\label{figfri}
Annual mean values of parameters used in Eqs.~(\ref{Wg}) and (\ref{Wg*}) to calculate global kinetic energy generation.
Empty red circles denote cell borders defined in (a) as the points where meridional velocity changes sign (see
also Fig.~\ref{cells}b).
}
\end{figure*}

To explore how expressions (\ref{Wg}) and (\ref{Wg*}) relate to observations we used NCAR/NCEP monthly re-analyses data
(averaged over the period from 1979-2013) on sea level pressure, geopotential height and air temperature at 13 pressure levels provided by the NOAA/OAR/ESRL PSD, Boulder, Colorado,
USA, from their Web site at http://www.esrl.noaa.gov/psd/ \citep{kalnay96}. All data were zonally averaged.
We defined the cell borders as positions where the zonally averaged meridional velocity at 1000~hPa changes its sign (Fig.~\ref{figfri}a). We
approximated $T_{ij}^-$ as the temperature of the top of the troposphere, having defined the latter, following
\citet{santer03}, as the height where the lapse rate diminishes to 2~K~km$^{-1}$.

In Fig.~\ref{figfri} we show the height of the troposphere $h_T$, cell borders, and temperatures for
the Northern and Southern hemispheres. We can see that $T^-$ is relatively constant across
the cells (Fig.~\ref{figfri}d), while the height of the troposphere (which is a proxy for cell height $z^-$)
decreases twofold within the Ferrel cells (Fig.~\ref{figfri}b).
This justifies the use of the Carnot formula for $A^-$, i.e. Eq.~(\ref{WK*}) rather than Eq.~(\ref{Wrr*})
for the horizontal streamline, to describe the upper streamline in the real cells.
The results are shown in Table~\ref{number}.

\begin{table*}
\scriptsize
\begin{minipage}[h]{0.99\textwidth}
\caption{\label{number}Budget of kinetic energy generation (J~mol$^{-1}$) calculated from Eqs.~(\ref{Wg}) and (\ref{Wg*})
for the heat engines (Hadley and Polar cells) and heat pumps (Ferrel cells) (Northern and Southern hemispheres combined)
for $z^+ = 1$~km and $c = 0.19$ ($\Gamma = 6.5$~K~km$^{-1}$). Uncertainty represents the standard deviation for monthly variation. "Total" in the last column includes the $(\Delta T_s/T^+)^2$ contribution from the upper atmosphere (this
term is absent in the lower atmosphere).
}
\begin{center}
\begin{tabular}{lrrrrrrrrrr}
\toprule
& \multicolumn{3}{c}{Lower atmosphere} & \multicolumn{4}{c}{Upper atmosphere} &	\multicolumn{3}{c}{Lower and upper atmosphere} \\
\cmidrule(r){2-4}\cmidrule(r){5-8}\cmidrule(r){9-11}
& Total &  $\Delta p_s/p$ & $\Delta T_s/T^+$ & Total & $\Delta p_s/p$ & $\Delta T_s/T^+$ &$(\Delta T_s/T^+)^2$ &  $\Delta p_s/p$ & $\Delta T_s/T^+$ &  Total \\
\cmidrule(r){2-4}\cmidrule(r){5-8}\cmidrule(r){9-11}
Ferrel &$44\pm 5$  &$30\pm 4$ &$14\pm 2$  &             $-152\pm 23$ &$-23\pm 3$&$-111\pm 17$&$-18\pm 4$&              $7\pm 2$       &$-97\pm 16$ &$-108\pm 18$\\
Hadley &$9\pm 1$   &$20\pm 2$ &$-11\pm 1$   &$108\pm 11$&$-15\pm 1$&$116\pm 11$           &$7\pm 1$&      $5 \pm 1$      &$105\pm 10$  &$117 \pm 11$\\
Polar  &$2\pm 2$   &$6\pm 3$        &$-4\pm 2$ &$19\pm 4$&$-5\pm 3 $&$18\pm 4$ &$6\pm 4$&          $0.7\pm 0.3$   &$14 \pm 4$   &$21 \pm 4$\\
All cells  &$55\pm 6$  &$56\pm 5$ &$-1\pm 3$  &           $-25\pm 23$&$-43\pm 3$ &  $23\pm 18$&$-5\pm 7$& $13\pm 2$&$22\pm 16$         &$30\pm 19$\\
\bottomrule
\end{tabular}
\end{center}
\end{minipage}
\end{table*}

For the lower atmosphere we find that total kinetic energy generation of $A_g^+ = 55 \pm 6$~J~mol$^{-1}$ is
determined by surface pressure differences, which make positive contributions to $A_g^+$ in all cells. The contribution of the temperature term
is virtually zero,
for $z^+ = 1$~km (boundary layer) in Eq.~(\ref{Wg}) it constitutes $-1\pm 3$~J~mol$^{-1}$.
The absolute magnitude of the contributions of $\Delta T_s$ within individual cells is substantially smaller than
that from $\Delta p_s$: it is positive in Ferrel cells and negative in Hadley cells,
contributing about 20\% of $A_g^+$ by absolute magnitude.

The mean kinetic energy generation by the upper atmosphere is negative, $A_g^- = -25 \pm 23$~J~mol$^{-1}$.
Surface pressure differences make negative contributions in all cells, while the surface temperature
contributions are positive in some cells, i.e. the Hadley and Polar cells as heat engines, and negative in
others, i.e. the two Ferrel cells as heat pumps. The linear temperature
term is also the most variable: this variability is both spatial and temporal reflecting seasonal
migration of cell borders.

Global kinetic energy generation is given by $A_g \equiv A_g^+ + A_g^- = 30\pm 19$~J~mol$^{-1}$,
which is 25\% of what is generated by the Hadley system ($A_{H} = 117 \pm 11$~J~mol$^{-1}$).
This is consistent with the analysis of \citet{kim13} who found that the net global meridional
kinetic power $W_{g}$ in NCAR-NCEP re-analysis is slightly positive, albeit
different methods of calculations gave different results: $W_g = 0.06$~W~m$^{-2}$ or $0.10$~W~m$^{-2}$.
Assuming that the Hadley system contributes $198\times 10^{12}$~W \citep{huang14} or $W_{H} = 0.4$~W~m$^{-2}$ globally,
we conclude that in the NCAR-NCEP re-analysis the net meridional kinetic power is 15-25\% of the
kinetic power in the Hadley system. This is in agreement with our results that are based
on NCAR-NCEP data: $W_g/W_H = A_{g}/A_{H}$ obtained from Eqs.~(\ref{Wg}) and (\ref{Wg*}) (Table~\ref{number}).
(Note that an equality between the ratios of power $W_i$ (W) and work $A_i$ (J~mol$^{-1}$) in different circulation systems implies an equality between
the amounts of gas $\mathcal{N}_i/\tau_i$ circulating per unit time along the considered
streamline in each system: $W_i = A_i \mathcal{N}_i/\tau_i$, where $\mathcal{N}_i$ (mol) is the
number of moles circulating along the considered streamline in the $i$-th circulation, $\tau_i$
is the time period of the cycle.)
In the MERRA re-analysis the net meridional kinetic power is slightly negative: $W_g = -0.06$~W~m$^{-2}$ or $-0.13$~W~m$^{-2}$ \citep{kim13}.
According to MERRA, the Ferrel cells consume more kinetic power than the Hadley cells produce \citep{huang14}.
The pattern common to both datasets is that most kinetic energy generated by the heat engines
is consumed by the heat pumps. This robust pattern is reproduced by our analysis (Table~\ref{number}).

\section{Discussion and conclusions}

\subsection{Kinetic energy generation as a function of surface pressure and temperature differences}

We derived how kinetic energy generation in the boundary layer, $A^+$ and $A_r^+$,
and in the upper atmosphere, $A^-$ and $A_r^-$, depends on surface pressure and temperature differences $\Delta p_s$ and $\Delta T_s$
that are measured along the air streamline.
This derivation is based on three fundamental relationships: the hydrostatic equilibrium, the ideal gas
law and the definition of mechanical work of an air parcel. The obtained expressions are valid for any air parcel
following a given trajectory irrespective of the angular velocity of planet rotation.

We applied the derived relationships to analyze kinetic energy generation in Earth's meridional
circulation cells.
On Earth a typical meridional surface temperature difference $\Delta T_s$ makes a larger
contribution to total kinetic energy generation $A$ than a typical surface pressure difference $\Delta p_s$.
E.g., for a Hadley cell $(1/c) \Delta T_s/T  \approx 10 \Delta p_s/p  \approx 0.1$.
Kinetic energy generation by a temperature-induced pressure gradient
in the upper atmosphere is by a similar factor larger than kinetic energy
generation by the surface pressure gradient at the lower isotherm, $A^- \gg A$, Eqs.~(\ref{WK})-(\ref{WK*}).
These relationships might have been responsible for the lack of explicit attention
to surface pressure gradients when considering the atmospheric thermodynamic cycles and kinetic energy budget.

However, as we have shown, the surface temperature gradient plays a two-sided role.
Firstly, it generates a pressure gradient in the upper atmosphere, which can be a source
of kinetic energy. Secondly, the same pressure gradient, if the air moves from the colder to the warmer region, consumes
kinetic energy. Such a "negative" cycle (heat pump) must be fed by a "positive" cycle (heat engine)
that has a positive work output.
Therefore, in the presence of a surface temperature gradient along which several circulation cells
are operating, the global efficiency of kinetic
energy generation can never reach Carnot efficiency: the circulation cells with negative work output can
reduce it to zero. A global Carnot efficiency for an atmosphere containing many circulation cells can be achieved only on
an isothermal surface where all cells are represented by Carnot cycles shown in Fig.~\ref{car}b ($\Delta p_s > 0$, $\Delta T_s = 0$, $A>0$).

\subsection{Small net kinetic energy generation in the upper atmosphere}

Our analytical formulations permit us to quantify the relative magnitudes of the surface pressure and
temperature contributions to kinetic energy generation.
We show that because of the large negative surface temperature differences across the Ferrel cells,
these cells consume most of the kinetic energy generated by the Hadley cells in the upper atmosphere (Table \ref{number}).
This is consistent with observations of the Ferrel and Hadley systems \citep[see][]{huang14,kim13}.

\begin{figure*}
\centerline{
\includegraphics[width=0.75\textwidth,angle=0,clip]{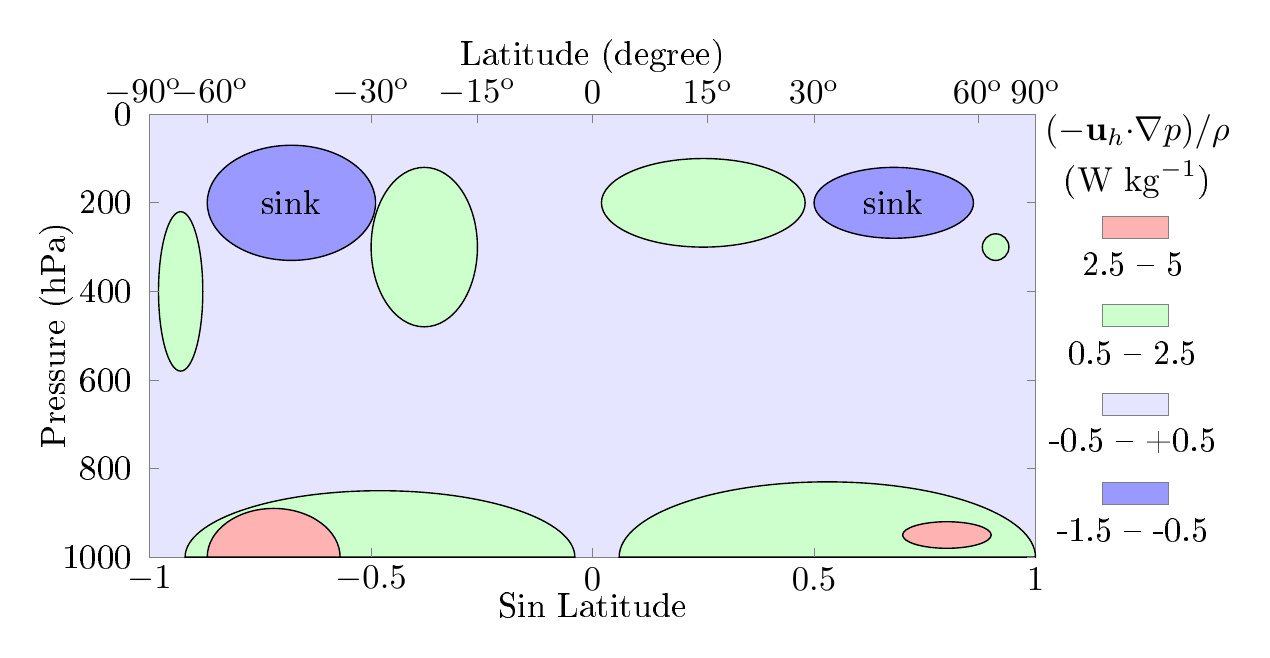}
}
\caption{\label{fighu}
Schematic representation of the budget of kinetic power generation in the upper and lower atmosphere redrawn
from Fig.~3a of \citet{huang15}. Note sinks of kinetic energy in the upper parts of the Ferrel cells.
}
\end{figure*}

Our simplified analysis does not permit us to estimate the net kinetic energy generation in the upper atmosphere with high precision.
However, our inspection of the up-to-date analysis of atmospheric kinetic power
reveals that the kinetic energy sources and sinks in the upper atmosphere virtually cancel each other out.
\citet{huang15} recently performed a high-resolution analysis of global kinetic power generation
using 3-hour $1.25\degree\times 1.25\degree$ gridded MERRA data (such an analysis obviously accounts for both
eddy and zonally averaged components of the circulation). These data, reproduced in Fig.~\ref{fighu},
show that in the lower atmosphere kinetic power is almost universally positive,
while in the upper atmosphere there are large sources and sinks of kinetic energy that are in approximate balance.
The net kinetic energy generation in the upper atmosphere appears to be small compared to the lower
atmosphere\footnote{We emphasize that the local rate of
kinetic energy generation in the upper atmosphere can be significantly
higher than it is on average in the lower atmosphere, cf.~Fig.~3b and 3c of \citet{huang15}.
Thus the small net value does not reflect a ubiquitous geostrophic balance (under geostrophic
balance kinetic energy generation would be zero everywhere).}.

To our knowledge, this remarkable pattern -- the cancellation of large energy sources and sinks
in the upper atmosphere -- has not been previously noted.
Negligible kinetic energy generation in the upper atmosphere means that the positive temperature contribution of circulation cells
that work as heat engines
must compensate the negative temperature contribution of all heat pumps as well as the negative pressure contribution
in both heat pumps and heat engines (Table~\ref{contrib}). Given that these contributions are governed by several independent
parameters, including surface pressure, temperature, lapse rate, height and width of the circulation cell, a random combination of these terms
resulting in a small net kinetic energy generation appears unlikely.
In our view, this compensation results from a dynamic constraint which
determines that the export/import of kinetic energy from/to the upper atmosphere is small. For a given
distribution of surface pressure and temperature, this condition can be satisfied by varying the height $z^-$
and temperature $T^-$ of the upper streamlines that determine kinetic energy
generation and its consumption in the upper troposphere (Table~\ref{contrib}). In other words, the observed height of the troposphere
as well as the isobaric height (\ref{ze}) can be determined by the condition that net kinetic energy generation
in the upper atmosphere is small. We suggested previously that rather than determining the ratio between surface pressure and temperature
differences \citep{lindzen87,bayr13}, these heights likely result from these differences \citep{jcli15}.

When kinetic energy generation in the upper atmosphere is negligible, global kinetic energy
generation is determined by surface pressure gradients. The temperature contribution in the lower atmosphere is relatively small due to the small
factor $z^+/h_s < 0.1$ (see Eqs.~\eqref{Wrze} and (\ref{AG})).

\subsection{The role of surface pressure gradients and condensation}

The obtained expressions for $A_r^+$ and $A^-$ show that in the upper atmosphere
kinetic energy generation grows with increasing $\Delta T_s$, while in the lower atmosphere it declines.
For a circulation cell where the air moves towards a warmer area, this means that
the larger the temperature difference between the warm and the cold ends of
the cell, the less kinetic energy is generated in the boundary layer.
This explains, for example, why maximum kinetic energy in the lower atmosphere is
generated within Ferrel cells, where $\Delta T_s$ is minimal (i.e. where $\Delta T_s < 0$
as the surface air moves towards a colder area, see Table~\ref{number} and Fig.~\ref{fighu}.)

As the pole is colder than the equator, the temperature difference $\Delta T_s$
across the meridional circulation cell grows with increasing meridional cell extension $L$.
The obtained theoretical expression for $A_r^+$ shows that if the surface pressure difference
$\Delta p_s$ remains constant, $A_r^+$ will decrease linearly with growing $L$. In such a case
the condition that $A_r^+$ in the boundary layer must be positive will limit the cell size.
Using monthly MERRA data, we showed that for the real-world Hadley cells $\Delta T_s$ does indeed grow with $L$,
but that $\Delta p_s$ reaches a plateau at some intermediate values of $L$ beyond which it does not further grow
with increasing $L$ despite increasing $\Delta T_s$. This means that there is a maximum value of kinetic energy
that can be generated in the lower atmosphere in Hadley cells (Fig.~\ref{figAr}).

The work performed by the turbulent friction force at the surface grows with increasing cell size
(and hence increasing streamline length). Since generation and dissipation of kinetic energy must balance each other,
this means that kinetic energy generation in the boundary layer cannot be less that a certain positive value corresponding to friction losses.
Other things being equal, if surface friction losses are reduced, the poleward cell extension can grow, an effect
noted by \citet{robinson97}. Meanwhile in the upper atmosphere, as noted in many studies \citep[e.g.,][]{held80,robinson97,marvel13},
friction has an opposite effect on cell size: the poleward cell extension grows with {\it increasing}
turbulent friction (eddy diffusivity) in the atmospheric interior (because of a smaller degree of momentum conservation
in the upper level flow as discussed in the Introduction).

Surface pressure gradients have been traditionally considered as outcomes of air re-distribution in the upper atmosphere caused by
temperature gradients \citep[e.g.,][]{pielke81}.
In terms of the energy budget of a thermodynamic cycle, this can be understood as follows.
A certain part of kinetic energy generated in the upper atmosphere in a heat engine is not consumed by heat pumps
but is converted to the potential energy of the surface pressure gradient. This potential energy
then is converted to the kinetic energy as the air moves in the boundary layer.
This explanation would be valid for the real Earth if we could show from theory that the upper-level air flow along the observed temperature
gradients do indeed generate enough potential energy to form a surface pressure gradient of observed
magnitude. However, since the upper level air flow and, hence, kinetic energy generation in the upper atmosphere,
can only be found by explicitly specifying turbulent friction, which is not known from theory, the question as to what
determines surface pressure gradients across the meridional cells remains open.
(We note that whatever is the cause of the surface pressure gradients apparently they do not always grow with
increasing $\Delta T_s$.)

As discussed in the introduction, the presence of the meridional temperature gradient alone is insufficient to generate a meridional
circulation of appreciable intensity and poleward extension. Some mechanism (like turbulence) must break the geostrophic equilibrium
and allow the upper-level air to actually move towards the pole. Only in such a case there appears a heat engine
that could generate some kinetic energy.
If we start from the observation that in the tropical upper troposphere there is a poleward air motion (which means that the geostrophic balance
is broken), we implicitly provide such a mechanism -- however without explaining its nature.
Then the properties of the low-level flow, including the surface pressure gradient, can be deduced and understood from the conservation
of mass and angular momentum as illustrated by both theory and observational analyses \citep[e.g.,][]{johnson89,cai14}. If, on the other hand,
we start from the observed pressure gradient and the intensity of air motion in the boundary layer, using the same conservation laws we can deduce
the upper-level air motions. However, as such diagnostic studies do not investigate the causes of the dynamic disequilibrium
that makes the meridional circulation possible and define its intensity in the lower and upper atmosphere, they do not identify the primary drivers of atmospheric circulation.

The explanation of surface pressure gradients as outcomes of temperature gradients is certainly
valid for a dry atmosphere where no other sources of potential energy for a surface pressure gradient can be thought of.
However, in a moist atmosphere there is a different process that directly impact surface pressure: condensation
and precipitation. In the Earth's atmosphere the instantaneous rates of condensation, which is proportional to vertical velocity,
can be two orders of magnitude greater than the instantaneous rates of evaporation.
Rapid removal of large amounts of water vapor necessarily disturbs whatever balance of forces
might have existed at the surface. The air converges towards this zone of low pressure.
Therefore, in a moist atmosphere that is unstable to condensation, a geostrophic equilibrium
and, thus, zero rate of kinetic energy generation are principally impossible.
If condensation continuously disturbs the geostrophic balance of a rotating atmosphere,
the power of the global atmospheric engine should be proportional to condensation rate
\citep[see also][]{acp13,jas13,pla14}.
Establishing a link between the wind power and condensation implies the need to revise
how turbulent friction is formulated in global circulation models.

Let us conclude by highlighting some implications of these mechanisms.
Kinetic power generation governed by the product of horizontal velocity and pressure gradient reflects
cross-isobaric motion towards the low pressure area and the associated air convergence. If
the kinetic power generation is proportional to condensation rate,
then over a dry continent where condensation is absent low-level air convergence will be strongly suppressed,
and a geostrophic (or cyclostrophic) balance will be established. The low pressure area over a dry hot land will not
lead to moisture convergence from the ocean, and drought will persist. This
indicates that removal of forest cover which is a significant store and source of moisture on land can lead to a self-perpetuating
drought. This mechanism may contribute to the recent major shifts in global rainfall patterns
like for example the recent catastrophic droughts in Brazil \citep{marengo15,dobrovolski15}.

We urge increased attention to the nature of surface pressure gradients on Earth and the role of condensation in their generation.

\section{Acknowledgment}

This work is partially supported by Russian Scientific Foundation Grant~14-22-00281, the University of California Agricultural Experiment Station, the Australian Research Council project DP160102107 and  the CNPq/CT-Hidro - GeoClima project Grant~404158/2013-7.

\section*{
\begin{center}
Appendix A. Work in the presence of phase transitions
\end{center}}
\label{apA}

Here we show that Eq.~(\ref{Wtotg}) accurately describes kinetic energy generation
even in the presence of phase transitions.
Consider an air parcel occupying volume $\tilde{V}\equiv \tilde{N} V$ and containing a total of $\tilde{N} = \tilde{N}_d + \tilde{N}_v$
moles of dry air and water vapor (subscript $d$ and $v$, respectively). This air parcel moves
along a closed stationary trajectory which can be described in $(p,V)$ coordinates (e.g.~Fig.~\ref{car}a).
Total work performed by such an air parcel per mole dry air is
\begin{align}
A_{p} &\equiv \frac{1}{\tilde{N}_d} \oint p d\tilde{V}  = \frac{1}{\tilde{N}_d} \left( \oint \tilde{N} p dV + \oint pV d\tilde{N} \right),
\quad \oint d\tilde{N} = 0.  \label{A1}
\end{align}
Here $d\tilde{N}=d\tilde{N}_v$ is the change of air amount due to phase transitions of water vapor.
Unlike $A$ in Eq.~(\ref{Wtotg}), $A_{p}$ (subscript $p$ refers to the presence of
phase transitions) is not a unique function of the integral
$\oint pdV$ that is determined by the area enclosed by the closed streamline in the $(p,V)$ diagram:
$A_{p}$ additionally depends on where along this trajectory the phase transitions take place.
This information is absent from the $p,V$ diagram.

Using the ideal gas law (\ref{ig}), the hydrostatic equilibrium (\ref{he}) and taking into account that $\oint d(\tilde{N}_v T) =
 \oint d(\tilde{N} T)= \tilde{N}_d \oint dT = 0$,
we find
\begin{align}
A_{p} =& -\frac{1}{\tilde{N_d}} \oint \tilde{N} Vdp  = -\frac{1}{\tilde{N_d}}\oint \tilde{N} V \left(\left.\frac{\partial p}{\partial y}\right|_{z=z(y)}dy
+ \left.\frac{\partial p}{\partial z}\right|_{y=y(z)}dz\right)   \label{A2} \\
=&-\oint (1 + \gamma_d) V \left.\frac{\partial p}{\partial y}\right|_{z=z(y)}dy + \frac{1}{\tilde{N_d}}\oint \tilde{N}M g dz. \label{A3}
\end{align}
Here $\gamma_d \equiv \tilde{N}_v/\tilde{N}_d\ll 1$. The second integral in Eq.~(\ref{A3}) reflects the difference between the air mass $\tilde{N} M$ (kg) that is rising ($dz > 0$) and descending ($dz < 0$) along the trajectory. This difference is caused by phase transitions (condensation and evaporation) that in the general case occur at different heights $z$. Therefore, the second integral in Eq.~(\ref{A3}) represents the difference in potential energy per mole dry air between the mean heights where condensation and evaporation take place. It is unrelated to the kinetic energy generation \citep[for further details see][]{pla15,he15}.

We conclude that in the presence of phase transitions, total work output $A_{p}$ is not equal to kinetic energy
generation because of the non-zero second integral in Eq.~(\ref{A3}), $A_{p} \ne A$.
However, the kinetic energy generation, which depends on the horizontal pressure gradient and is described by the first integral in Eq.~(\ref{A3}),
coincides with $A$ (\ref{Wtotg}) with good accuracy because of the small value of $\gamma_d \ll 1$.

\begin{figure*}
\centerline{
\includegraphics[width=0.95\textwidth,angle=0,clip]{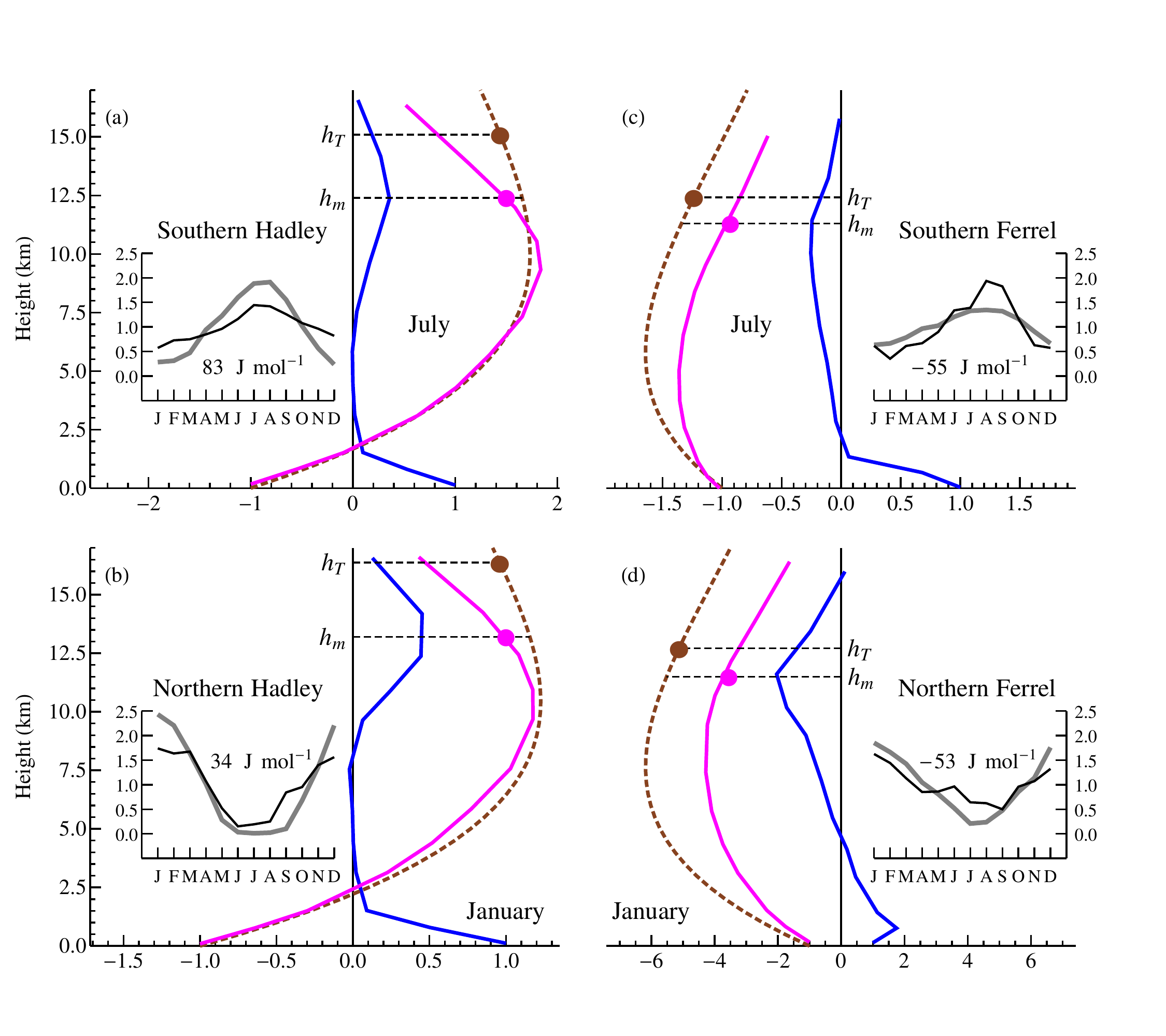}
}
\caption{\label{diff}
Vertical profiles of the observed (solid pink) and theoretical (dashed brown) pressure
differences $\Delta p(z)$ (cf. Fig.~\ref{car}d) between the borders of (a) Southern Hadley cell in July, (b) Northern Hadley cell in January,  (c) Southern Ferrel cell in July, (d) Northern Ferrel cell in January. During these months
the cells have the maximum power output (see the insets, see key below).
The blue curves indicate estimated kinetic power generation: the product of the mean meridional pressure gradient within the cell
by the mean meridional velocity as dependent on height $z$ as a proxy measure. All vertical profiles are normalized
by their value at the surface. The small filled circles denote the theoretical pressure difference
at the top of the troposphere $z = h_T$ (mean $h_T$ within the cell is shown) and the observed
pressure difference at $z = h_m$, the height where kinetic energy generation in the upper troposphere is maximum.
The data are long-term mean NCEP-NCAR climatology (see Section~\ref{hp}).
The insets show the monthly variation of the theoretical work output for each cell (thin black lines), Eqs.~(\ref{Wg}), (\ref{Wg*}),
compared to the monthly variation of the observed power output in the same cell (thick gray lines) according to the data
of Fig.~2 and 7 of \citet{huang14}. The monthly values are normalized by the annual mean (for the work output the annual mean value
is shown on the graph).
}
\end{figure*}

\section*{
\begin{center}
Appendix B. Validity of the theoretical approach
\end{center}}
\label{apB}

Our derivations have assumed an atmosphere with a constant lapse rate and idealized air trajectories.
We now examine how these assumptions impact our two major findings:
first, how maximum cell size depends on surface pressure and temperature differences,
Eqs.~(\ref{cond}) and (\ref{ze}); second, our conclusion that in the meridional cells the balance between kinetic energy
generated by heat engines and consumed by heat pumps reflects the meridional differences in surface
temperature (Table~\ref{contrib} and \ref{number}).

The first finding is based on the expression for kinetic energy generation in the lower atmosphere:
$A^+$ (Eq.~\eqref{WK}) for Carnot and $A_r^+$ (Eq.~\eqref{Wrr}) for the rectangular cycle.
Since $A_r^+$
is linear over $\Delta p_s$, $\Delta T_s$ and $z$ (height of air motion), Eq.~(\ref{Wrr}) with $z = \overline{z}$ can be applied to any
trajectory of air motion with a sufficiently small mean height $\overline{z}$.
This is because even if the lapse rate $\Gamma$ varies in the horizontal and/or vertical, the smallness of $\overline{z}$
will ensure that $T(\overline{z}) \approx T^+$ (air temperature
is approximately equal to surface temperature) in the pressure term in Eq.~(\ref{Wrr}).
For $z = 1$~km and a typical tropical ratio $\Delta p_s/p \sim (1/3)\Delta T_s/T$ \citep{lindzen87,bayr13,jcli15}, the pressure
term is the main one determining the value of $A_r^+$ (\ref{Wrr}).
In \ref{apC} we discuss that any possible impact of the spatial variation in lapse rate
is also negligible.

As we show in Fig.~\ref{diff}, most kinetic energy in the lower half of the troposphere is generated within a narrow
boundary layer: the rate of kinetic energy generation diminishes linearly with increasing altitude approaching zero for $z \approx 2$~km
(the Northern Ferrel cell is an exception discussed below).
If within this layer the distribution of air pressure
is satisfactorily described by Eq.~(\ref{p}) with a constant lapse rate, then our formula for kinetic energy generation in the lower atmosphere
$A_r^+$, Eq.~(\ref{Wrr}), is valid (in calculations in Table~\ref{number} we used $z = 1$~km).
Indeed, Fig.~\ref{diff}a,b shows that in the Hadley cells in the lowest 2~km the observed pressure
difference across the cell is very close to the theoretical pressure difference calculated
from Eq.~(\ref{p}). The reason is that the pressure scale height $h_s$ (\ref{Wrze}) is governed by surface temperature,
such that whatever the differences there are in the lapse rates, given $z$ is small, they cannot
significantly change this basic height in the boundary layer. Any effect of the actual form of the air trajectory
or of variations in lapse rate in time and space will be negligible.
We thus conclude that Eqs.~(\ref{Wrr}) and (\ref{ze}) for the rectangular cycle are always valid for
the Hadley cells.

Note that in the Northern Ferrel cell the generation of kinetic energy is not confined to the lower 2~km
of the atmosphere (presumably because this cell harbors
most land including mountains) (Fig.~\ref{diff}c).
In this cell our theoretical relationship apparently overestimates the observed pressure difference already in the lower troposphere.
However, the two inaccuracies partially cancel each other in our estimate in Table~\ref{number}:
the mean height where kinetic energy is generated in the lower part of the Northern Ferrel cell is
approximately twice the value of $z = 1$~km that we assumed for all cells;
but the observed pressure difference at this height is about 1.5 times lower than the theoretical pressure difference (Fig.~\ref{diff}c).

The relationships for the Carnot cycle in the lower atmosphere, Eqs.~(\ref{WK}) and (\ref{cond}), presume that the mean height of the lower streamline
is proportional to $\Delta T_s$. In the real cells the mean height of the low-level air trajectory does not vary linearly with $\Delta T_s$.
The Carnot relationships have a theoretical value: they explain how because of the heat pumps an ideal Carnot efficiency in the atmosphere as a whole
can be inachievable in the presence of meridional gradients of surface temperature.
While the presumably low efficiency of the atmospheric engine
has been blamed on the hydrological cycle \citep[e.g.,][]{pauluis11}, our analysis suggests a different source of inefficiencies.

In the upper atmosphere the discrepancy between the theoretical and observed pressure distributions
is apparently more significant (Fig.~\ref{diff}): over this larger range in altitude the lapse rate variation finally plays in.
However, to justify our main conclusion --
highlighting cancellation of kinetic energy
among cells through processes primarily related to surface temperature -- it is sufficient to capture the horizontal
pressure difference at the height $h_m$ where kinetic energy generation is maximum.

For the Hadley cells, the theoretical and observed pressure difference remain close from $z = 0$ up to
$z \approx h_m$. For $z > h_m$ the observed pressure difference declines more rapidly than does the theoretical one.
For this reason, while the estimated top of the troposphere $h_T$ is located higher than $h_m$, the
theoretical pressure difference estimated by us for $z = h_T$ (Table~\ref{number}) is
very similar to the observed pressure difference at $h_m$ (Fig.~\ref{diff}).
In the Ferrel cells $h_T$ and $h_m$ approximately coincide, while the discrepancy between the theoretical and observed pressure distribution is
larger than in the Hadley cells but never exceeds 30\%.
Overall, Fig.~\ref{diff} makes it clear that Eqs.~(\ref{Wg}) and (\ref{Wg*}), which focus
on surface temperature and pressure differences, satisfactorily capture the behavior of the pressure difference
across the circulation cells and thus provide a reasonable first-order estimate for the relationship
between work outputs of the Earth's major heat engines and heat pumps.

\section*{
\begin{center}
Appendix C. Spatial variation of lapse rate in the lower atmosphere
\end{center}}
\label{apC}

Here we show that Eq.~(\ref{Wrze}) remains valid in the presence
of spatial variation in temperature lapse rate $\Gamma\equiv -\partial T/\partial z$ if
under $\Delta T_s$ in Eq.~(\ref{Wrze}) we understand the horizontal temperature difference
at one half the mean height of the lower streamline. In the formulae below $T \equiv T(y,z) = T_s(y) - \Gamma(y) z(y)$,
$p \equiv p(y,z) = p_s(y) [T(y,z)/T_s(y)]^{1/c(y)}$,
where $T_s(y)$ is surface temperature and $p_s(y)$ is surface pressure.

Using these relationships we find
\begin{align}
&\left.\frac{RT}{p}\frac{\partial p}{\partial y} \right|_{z=z(y)} =
\frac{RT}{p_s}\frac{\partial p_s}{\partial y}  + \Psi,  \nonumber\\
&\Psi\equiv
\frac{RT}{c} \left(\frac{1}{T}\frac{\partial T}{\partial y}  -
\frac{1}{T_s} \frac{\partial T_s}{\partial y} - \frac{1}{c} \frac{\partial c}{\partial y} \ln \frac{T}{T_s}\right).  \label{B1}
\end{align}
Expanding the logarithm in Eq.~(\ref{B1}) over $\Gamma z/T_s \ll 1$
\begin{equation}\label{B2}
\ln \frac{T}{T_s} = \ln \left(1 - \frac{\Gamma z}{T_s}\right) = - \frac{\Gamma z}{T_s} - \frac{1}{2}\left(\frac{\Gamma z}{T_s}\right)^2,
\end{equation}
we find from Eq.~(\ref{B2}) (note that $R/c = M g/\Gamma$ and $T/T_s = 1 - \Gamma z/T_s$)
\begin{align}
\label{B3}
\Psi = M g z\left(\frac{1}{T_s}\frac{\partial T_s}{\partial y} - \frac{z}{2} \frac{1}{T_s}\frac{\partial \Gamma}{\partial y} -\frac{1}{z}\frac{\partial z}{\partial y} \right).
\end{align}
When $\Psi$ is integrated over the boundary layer of a fixed height, i.e. when $z(y_1)=z(y_2)$, the last term in Eq.~(\ref{B3}) vanishes.

\begin{figure*}
\begin{minipage}[h]{0.99\textwidth}
\centerline{
\includegraphics[width=0.99\textwidth,angle=0,clip]{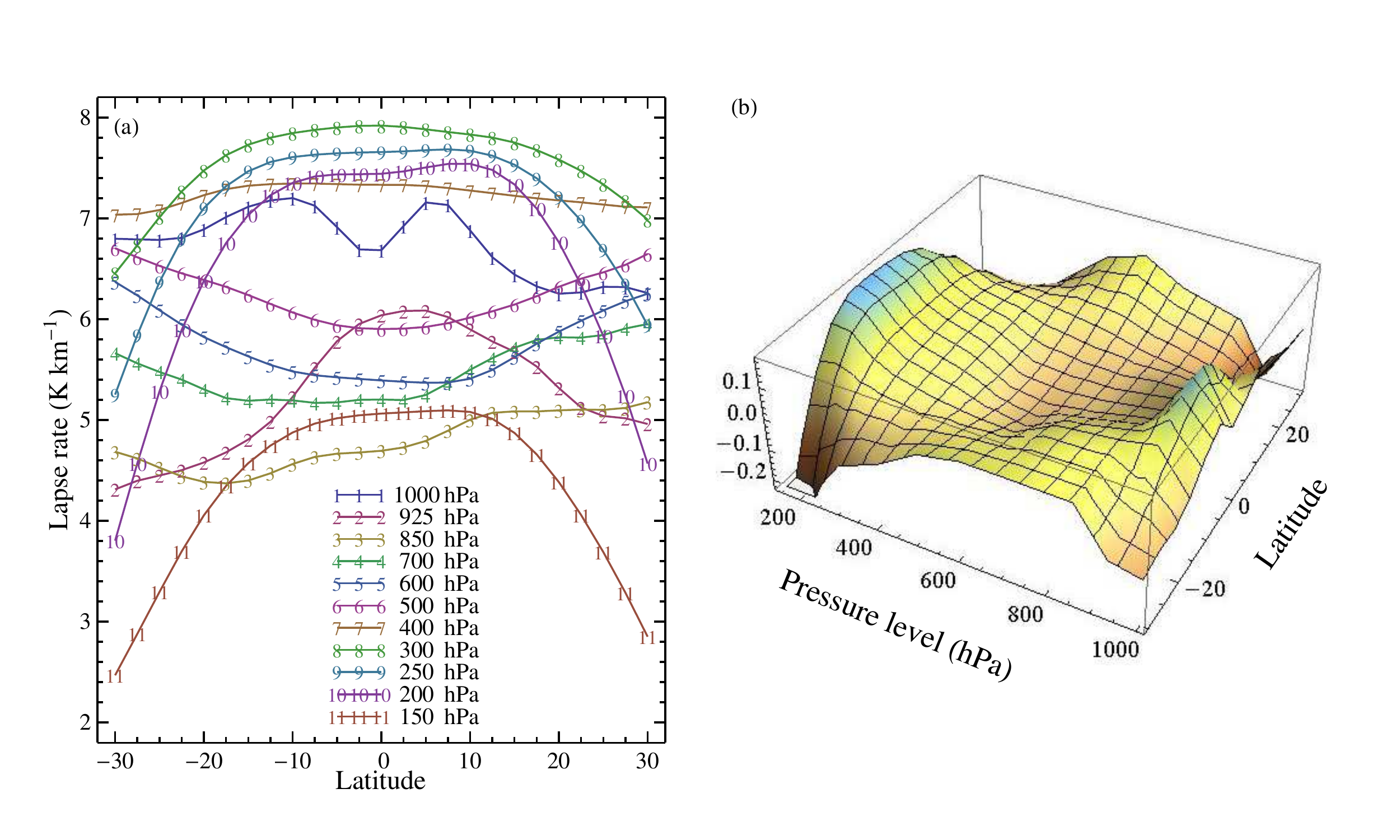}
}
\end{minipage}
\caption{\label{figlap}
Annual mean latitudinal profiles of the air temperature lapse rate on different
pressure levels. For example, curve 1 in (a) shows the mean lapse rate between 1000 hPa
and 925 hPa; curve 2 -- between 925 hPa and 850 hPa; curve 11 -- between 150 and 100 hPa.
Panel (b) shows the relative horizontal variation
-- at each pressure level the lapse rate at a given latitude is divided by the mean lapse rate at this level (averaged between
30$\degree$S and 30$\degree$N). The equator has a higher lapse rate than the 30th latitudes in the lower and upper
-- but not the middle -- troposphere. The data are long-term mean NCEP-NCAR climatology (see Section~\ref{hp}).
}
\end{figure*}

Noting that $\Gamma z \ll T$, $T \approx T_s$ and $\Delta p_s \ll p_s$
and that we defined $\Delta p_s \equiv -\int_{y_1}^{y_2} (\partial p_s/\partial y) dy$,
the kinetic energy generation in the presence of lapse rate variation  $A_\Gamma$
becomes, cf. Eq.~(\ref{Wrze}):
\begin{align}
A_\Gamma &\equiv -\int_{y_1}^{y_2} \left.\frac{RT}{p}\frac{\partial p}{\partial y} \right|_{z=z(y)} dy \nonumber\\
&\approx  Mg h_s \left[\frac{\Delta p_s}{p_s} - \frac{\overline{z}}{h_s} \frac{\Delta T}{T_s}\right],  \label{AG}\\
\Delta T &\equiv \Delta T_s - \Delta \Gamma \frac{\overline{z}}{2}.   \label{dT}
\end{align}
Here $\Delta T$ is the horizontal temperature difference at a height equal to one half of the mean streamline height $\overline{z}$.

Equation~(\ref{AG}) allows us to conclude that with a small $\overline{z} \sim 1$~km any horizontal variation in lapse rate
makes a minor contribution to kinetic energy generation. (The vertical variation in $\Gamma$ is zero in the first approximation,
again because $z$ is small). For example, even if in the boundary layer the lapse rate changes by $\Delta \Gamma = 4$~K~km$^{-1}$
(which is approximately the difference
between the dry and moist adiabatic lapse rates), the lapse rate term in Eq.~(\ref{AG}) will be negligible
compared to the pressure term: $(\overline{z}/h_s) (\Delta \Gamma \overline{z})/(2 T_s) \sim  0.7 \times 10^{-4} \ll  \Delta p_s/p_s \sim 10^{-2}$.

Note that in the real atmosphere the horizontal variation in lapse rate is smaller.
For example, the zonally averaged lapse rate between 925 and 850 hPa changes across the Hadley cells by about $2$~K~km$^{-1}$ (Fig.~\ref{figlap}).
Notably, in the lower atmosphere the equatorial areas of the Hadley cells have a larger lapse rate than
the higher latitudes ($\Delta \Gamma > 0$). This lapse rate variation makes a positive contribution
to the kinetic energy generation similar to the positive contribution of the negative surface temperature gradient in
the Ferrel cells.

In severe hurricanes the horizontal variation in lapse rate across the hurricane area can be about
$\Delta \Gamma = -4$~K~km$^{-1}$ \citep[e.g.,][their~Fig.~4c]{montgomery06} in the lower 2~km. This variation makes a negative contribution
to kinetic energy generation. But
in these circulation systems $\Delta p_s/p_s > 5\times 10^{-2}$ is several times greater than in the zonally averaged cells,
so the pressure term in Eq.~(\ref{AG}) remains the main one. These considerations indicate that in the lower atmosphere kinetic energy generation is
determined by surface pressure gradients.


\end{document}